\documentclass{aa}  

\usepackage{graphicx}
\usepackage{xcolor}
\usepackage{txfonts}

\newcommand{\Teff}{\mbox{$T_\mathrm{eff}$}\xspace}

\newcommand{\Lsol}{$\mathrm{L}_\odot$}
\newcommand{\Msol}{$\mathrm{M}_\odot$}
\newcommand{\Rsol}{$\mathrm{R}_\odot$}

\usepackage{hyperref}
\hypersetup{
    colorlinks=true,
    linkcolor=cyan,
    citecolor=cyan,
    urlcolor=cyan,
    }

\newcommand{\orcid}[1]{\protect\href{https://orcid.org/#1}{\protect\includegraphics[width=8pt]{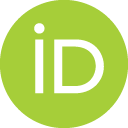}}}
\begin{document}

   \title{K 1-6 is a photoionised ISM nebula shaped by a fast-moving hot white dwarf in a triple system}

   \subtitle{}

   \author{Jaroslav Merc\orcid{0000-0001-6355-2468}\inst{1,2}
          \and
          David Jones\orcid{0000-0003-3947-5946}\inst{2,3}
          \and
          Ana Escorza\orcid{0000-0003-3833-2513}\inst{2,3}
          \and
          Henri M. J.\ Boffin\orcid{0000-0002-9486-4840}\inst{4}
          \and
          Nicole Reindl\orcid{0000-0002-0119-7883}\inst{5}
          \and
           Michael~Abdul-Masih\orcid{0000-0001-6566-7568}\inst{2,3}
          \and
           Thomas~Masseron\orcid{0000-0002-6939-0831}\inst{2,3}
          \and
          Paulina~Sowicka\orcid{0000-0002-6605-0268}\inst{2,3}
          \and
          Jan Kára\orcid{0000-0002-1012-7203}\inst{6}
          \and
          Jorge~García-Rojas\orcid{0000-0002-6138-1869}\inst{2,3}
          \and
          Rafael~A.~García\orcid{0000-0002-8854-3776}\inst{7}
          \and
        Dinil~B.~Palakkatharappil\orcid{0000-0002-6812-4443}\inst{7}
          \and
          Quentin A. Parker\orcid{0000-0002-2062-0173}\inst{8}
          \and
          Albert A. Zijlstra\orcid{0000-0002-3171-5469}\inst{9}
          \and
          Diego~Godoy-Rivera\orcid{0000-0003-4556-1277}\inst{2,3}
          \and
          Paul~G.~Beck\orcid{0000-0003-4745-2242}\inst{2,3}
          \and
          Savita~Mathur\orcid{0000-0002-0129-0316}\inst{2,3}
          \and
          Montserrat~Armas~Padilla\orcid{0000-0002-4344-7334}\inst{2,3}
          \and
          Teo~Muñoz~Darias\orcid{0000-0002-3348-4035}\inst{2,3}
          \and
          Hana~Kučáková\orcid{0000-0002-1330-1318}\inst{1,10,11}
          \and
          Marek~Wolf\orcid{0000-0002-4387-6358}\inst{1}
          \and
          Kamil~Hornoch\orcid{0000-0002-0835-225X}\inst{11}
          \and
          Petr~Zasche\orcid{0000-0001-9383-7704}\inst{1}
          \and
          Rosa~Clavero\orcid{0000-0002-5435-0634}\inst{2,3}
          \and
          Peter~Goodhew\orcid{0009-0005-3715-4374}\inst{12}
          }

    \authorrunning{Jaroslav Merc et al.}
    \titlerunning{Photoionised ISM nebula in the triple system K 1-6}

   \institute{Astronomical Institute of Charles University, V Hole\v{s}ovi\v{c}k{\'a}ch 2, Prague, 18000, Czech Republic\\
              \email{jaroslav.merc@mff.cuni.cz}
        \and
        Instituto de Astrof\'isica de Canarias, Calle Vía Láctea, s/n, E-38205 La Laguna, Tenerife, Spain
        \and
        Departamento de Astrof\'isica, Universidad de La Laguna, E-38206 La Laguna, Tenerife, Spain
        \and
        European Southern Observatory, Karl-Schwarzschild-str. 2, D-85748 Garching, Germany
        \and
        Landessternwarte Heidelberg, Zentrum für Astronomie, Ruprecht-Karls-Universität, Königstuhl 12, 69117, Heidelberg, Germany
        \and
        Department of Physics and Astronomy, University of Texas Rio Grande Valley, Brownsville, TX, 78520, USA
        \and
        Universit\'e Paris-Saclay, Universit\'e Paris Cit\'e, CEA, CNRS, AIM, 91191, Gif-sur-Yvette, France
        \and
        The Laboratory for Space Research, The University of Hong Kong, Cyberport 4, Pokfulam, Hong Kong, People's Republic of China
        \and
        Jodrell Bank Centre for Astrophysics, Department of Physics and Astronomy, The University of Manchester, Oxford Road, Manchester M13 9PL, UK
        \and
        Research Centre for Theoretical Physics and Astrophysics, Institute of Physics, Silesian University in Opava, Bezručovo nám. 13, 746 01 Opava, Czech Republic
        \and
        Astronomical Institute of the Czech Academy of Sciences, Fričova 298, CZ-251 65 Ondřejov, Czech Republic
        \and
        Deep Space Imaging Network, 108 Sutton Court Road, London, W4 3EQ, UK
        }

   \date{Received \today; accepted XXX}

 
  \abstract
   {}
   {K 1-6 has long been classified as a planetary nebula (PN) hosting a binary central star, yet it has remained poorly studied due to its faintness. The central star exhibits pronounced photometric variability whose origin has so far been unclear. We aim to present a comprehensive characterisation of the K 1-6 system, including the physical properties of its stellar components and the nature of the surrounding nebulosity.}
   {We conducted a multi-wavelength analysis combining optical and UV spectroscopy obtained with the Gran Telescopio Canarias, the Telescopio Nazionale Galileo, the Nordic Optical Telescope, and the Hubble Space Telescope. We also present long-term multi-band ground- and space-based photometry, including high-cadence data from the Transiting Exoplanet Survey Satellite, narrow-band imaging, and the latest astrometric constraints from \textit{Gaia}. }
   {Our results show that the nebula is not a remnant PN, but instead consists of interstellar medium photoionised by a hot white dwarf, which is relatively evolved. It has a cooling age of 1–2 Myr, implying that any original PN has long since dissipated. We further find that the central object is a hierarchical triple system, comprising an inner binary with an orbital period likely of the order of thousands of days and a distant tertiary companion on a timescale of tens of thousands of years. The optically dominant cool component of the inner binary is an inflated K-type star displaying extreme magnetic activity, including large-amplitude variability and flaring. Its properties resemble those of BY Dra-type binaries and Abell 35-type systems, and are difficult to reconcile with single-star evolution, pointing instead to a history of binary interaction.}
   {}


   \maketitle
%

\section{Introduction}

The nebula Kohoutek~1-6 (PN~K1-6, central star V434 Dra/TIC 237292969; $\alpha_{\rm 2000}$ = 20:04:14.28, $\delta_{\rm 2000}$ = +74:25:35.93) was discovered on Palomar Observatory Sky Survey data and classified as a planetary nebula (PN) candidate by the Czech astronomer \citet{1962BAICz..13..120K}. He noted the detection of an almost circular disc of low density, more clearly visible in the red plates. Since its discovery, K~1-6 has not been studied in much detail, given its faintness \citep[e.g.,][]{1968BAICz..19...90K}. \citet{1988AJ.....96..997K} failed to detect the central star (CSPN) of K~1-6, and as a result, its classification as a PN nature remained uncertain for some time. The photometric variability of the star, now associated with K~1-6, was first discovered thanks to the Northern Sky Variability Survey (NSVS) data \citep[NSVS 1225362;][]{2008OEJV...88....1U}. The authors classified the star as an RS CVn variable, whereas \citet{2009AJ....138..466H} later classified it as a possible Cepheid. 

The photometric variability, the appearance of the nebula, and its nature were studied in greater detail by \cite{frew11}. Thanks to the \textit{GALEX} satellite ultraviolet (UV) observations, they identified the hot CSPN (a hot sub-dwarf or white dwarf) and argued that the nebula is a true PN. Based on the multi-frequency spectral energy distribution (SED), \cite{frew11} suggested that the optical and infrared part of the spectrum is dominated by a cool star, which they classified as a late G or early K-type subgiant or giant. Lacking spectroscopic data, its luminosity class remained uncertain. Nevertheless, this established K 1-6 as a PN whose CSPN is at least a binary system. The authors also confirmed the photometric variability of the CSPN with a period of 21.3 days. The most likely explanation for the photometric changes is RS CVn variability (and, in that case, they speculated that the CSPN may be ternary, as most RS CVn systems contain two cool stars), although an FK Com classification was also possible. The detection of the star by the \textit{ROSAT} satellite \citep{1999A&A...349..389V} in X-rays provides good support for the idea that the cool star is chromospherically active and that the 21.3-day period is likely rotational. Narrow-band H$\alpha$ + [\ion{N}{ii}] and [\ion{O}{iii}] imaging also revealed clear interaction of the nebula with the surrounding interstellar medium (ISM).

The apparent presence of both a hot ionising source and an active cool companion naturally raised the question of whether K~1-6 might belong to the growing class of PNe with binary or even higher-order multiple central stars. Binarity is now recognised as a key factor in shaping PN morphologies and in explaining the diverse properties of their central stars \citep[e.g.,][]{2017NatAs...1E.117J,2019ibfe.book.....B}. While close post-common-envelope binaries are well documented, {systems containing additional, wider companions remain rare, and the physical impact of these companions on the central star evolution is often debated. However, wider binary central stars are less commonly observed but not absent \citep[see, e.g.,][]{2017A&A...600L...9J,2022MNRAS.509.2566M,2025ApJ...980..227C}}. The complex photometric behaviour and the interacting-nebula morphology of K 1-6 therefore make it a promising candidate for a more intricate, possibly higher-order multiple system.

Triple stellar evolution has been proposed as a mechanism for producing the diverse and often asymmetric morphologies of many PNe \citep{bear17}, though direct evidence for triple central stars remains scarce. {The central star of NGC 246 is part of a common proper motion triple \citep{adam14} with a separation of about 500 au. Although such wide tertiaries were traditionally considered dynamically unimportant, recent theoretical work suggests that hierarchical triples can promote mass transfer in wide binaries through secular dynamical evolution, making them potentially important for the formation of post-mass-transfer WD+main-sequence systems \citep[see, e.g., ][and the Appendix A of \citealt{2026arXiv260323756B}]{2016ComAC...3....6T,2020A&A...640A..16T,2022A&A...661A..61T,2022MNRAS.514.1895K,2023ApJ...950....9R,2025A&A...704A.156R,2023ApJ...955L..14S,2025PASP..137i4201S,2025ApJ...978...47S,2026enap....2..279P}.} Additional examples of potential triples among central stars of PNe include the central binary star of DS~1, KV Vel, that has been shown to present very small cyclical variations in the timing of its photometric minima, which could be consistent with a brown dwarf tertiary companion \citep{qian24}. \citet{2024MNRAS.528.3392W} suggested that the Ring Nebula (NGC 6720) might also host a triple star system based on the regularly spaced concentric arc-like features in JWST observations, suggesting a low-mass companion with an orbital period of 280 years, in addition to a previously known wide M2-4 dwarf companion at a projected separation of 15\,000 au. Detailed study of NGC~3132 by \citet{demarco22} indicated that its central star may be part of a trinary or even quaternary system, but the presence of the third and fourth companions is speculatively inferred based on the nebular morphology (as such, it may only be a relatively wide binary); see also the analysis by \citep{2023ApJ...943..110S} who reach a similar conclusion. Sp~3 has also been suggested to be a triple, but the primary argument for its classification is the erroneous inversion of a negative parallax \citep{miszalski19}. Sh~2-71 has been proposed as a triple system that broke up during the formation of the PN \citep{jones19}. Finally, \citet{ciardullo99} identified a potential tertiary component (based on its projected proximity) to the binary central star of A~63, but this has yet to be confirmed via proper motions or radial velocities (we, in fact, reject this hypothesis based on our analysis presented in Appendix \ref{app:a63}).

On top of the possible multiple-star nature of K~1-6, \citet{2013MNRAS.436.2082T} suggested, based on the results of \citet{frew11}, that the object could be analogous to systems such as LoTr~1 and related PNe hosting s-process–enriched cool central stars within ring-like nebulae. However, no stellar spectroscopy was available at that time to confirm this hypothesis. The first low-resolution spectrum of the central star of K~1-6 was later obtained by \citet{2021MNRAS.506.4151M} as part of a survey of symbiotic-star candidates. Combining these observations with the SED, infrared colours, and \textit{Gaia} EDR3 data, the authors concluded that the cool component is a K1--3-type main-sequence star.

To complicate the picture even further, more recently, \citet{2025MNRAS.543.3035C} included K~1-6 in their study of extinction toward PN central stars and, intriguingly, questioned its PN classification. They pointed out the unusually low luminosity of the supposed ionising source and several peculiar properties of the nebula, issues that we revisit in this work using new multi-wavelength observations and a comprehensive re-analysis of the system.

In this paper, we present new imaging, photometric, and spectroscopic observations of K~1-6, including ultraviolet data from the Hubble Space Telescope (HST). These data allow us to re-examine both the nature of the nebula and the properties of its central star. We show that the nebula is not a true PN but rather an interstellar cloud photoionised by a hot white dwarf (WD). Our analysis further reveals that the central object is a complex, triple system, in which the dominant optical component is a highly active K-type inflated main-sequence star exhibiting long-lived spotted variability.

\section{Observations}

\subsection{Photometry}
Photometric monitoring of K~1-6 was carried out on 23 individual nights between July 2020 and November 2021 using the 65-cm Mayer Telescope (D65) at the Ondřejov Observatory, equipped with a Moravian Instruments CCD camera (MII G2-3200) and Johnson-Cousin \textit{BVRI} filters. Additional $V$- and $g$-band data were obtained from the All-Sky Automated Survey for Supernovae \citep[ASAS-SN;][]{2014ApJ...788...48S,2017PASP..129j4502K} and cover the interval between June 2012 and May 2025. Further $g$-band measurements, obtained between March 2018 and October 2024, are available from the Zwicky Transient Facility \citep[ZTF;][]{2019PASP..131a8003M}. In addition, $o$- and $c$-band photometry from the Asteroid Terrestrial-impact Last Alert System \citep[ATLAS;][]{2018PASP..130f4505T,2020PASP..132h5002S}, covering the interval from July 2015 to May 2025, were obtained via the ATLAS Forced Photometry server \citep{2021TNSAN...7....1S}.

K~1-6 was also observed by the \textit{Transiting Exoplanet Survey Satellite} \citep[TESS;][]{2015JATIS...1a4003R}, with data available from 31 sectors spanning the period from September 2019 (Sector 16) to December 2024 (Sector 86). These observations, though interrupted by several gaps lasting up to a few months, provide three extended light curves. Full-frame images (FFIs) are available at a 30-min cadence for Sectors 16--26, at a 10-min cadence up to Sector 55, and subsequently at a 200-s cadence. The analysis was performed using light curves produced by the Quick-Look Pipeline \citep[QLP;][]{2020RNAAS...4..204H}, which provides systematic reductions of TESS FFIs. In these data, the Simple Aperture Photometry (SAP) flux for each sector is independently normalised, so the light curves from different sectors had to be combined into a continuous time series. This was achieved by folding the light curve over a range of trial periods and adjusting the flux levels of individual sectors to minimise scatter in the phase-folded curve. For each trial period, the standard deviation of the residuals relative to the phase-binned mean was calculated. The period that minimised this residual scatter was adopted as the best-fitting period.

\subsection{Spectroscopy}
To investigate potential radial velocity variability and estimate the parameters of the optically dominant star, we observed the central star of K~1-6 using the FIES spectrograph \citep{fies} mounted on the 2.56-m Nordic Optical Telescope \citep[NOT;][]{not}. Observations were carried out in high-resolution mode, with a fibre diameter of 1.3\arcsec{}, yielding a reciprocal resolution of approximately 67\,000 over the wavelength range 3\,700--9\,000~\AA{}. All data were reduced using the standard FIEStool pipeline\footnote{\url{http://www.not.iac.es/instruments/fies/fiestool/FIEStool.html}}. In total, 17 spectra were obtained over 11 observing nights between March 4, 2023, and 
February 11, 2026 (the final epoch was observed under program 70-NOT7/25B (PI: Sowicka), while the rest were obtained in staff time). Two of these spectra, affected by very low signal-to-noise ratios, were excluded from the analysis.

Additionally, we obtained two spectra of the central star of K~1-6 with the HARPS-N spectrograph \citep{HarpsN} on the 3.57-m Telescopio Nazionale Galileo (TNG) to extend the time baseline. The observations were carried out on June 17, 2025 and September 23, 2025 under program SST2025-695 (PI: Escorza). 
HARPS-N covers the wavelength range 3\,830–6\,930 \AA{} and, with its 1\arcsec{} fibre, provides an average resolution of about 115\,000. The exposure time was 3\,000~s, and the spectrum was reduced using the standard HARPS-N Data Reduction Software \citep{HarpsN-DRS}.

The nebular spectrum analysed in this study was obtained with the OSIRIS long-slit spectrograph on the 10.4-m Gran Telescopio Canarias (GTC), as part of a broader survey by \citet{2023MNRAS.520..773R}. The observation was conducted on May 16, 2016, and the spectrum is publicly accessible through the GTC Public Archive\footnote{\url{https://gtc.sdc.cab.inta-csic.es/gtc/}}. Details of the data reduction can be found in Sect. 3 of \citet{2023MNRAS.520..773R}.

The central star of K~1-6 was also observed for 2378\,s on February 26, 2023, with the Cosmic Origins Spectrograph (COS) aboard the HST, using the G130M grating centred on 1291\AA\ (dataset LEZCI2020, PI: Reindl). On the same day, another 1537\,s exposure was taken with HST/COS using the G140L grating ($R\approx 1000$) that covers $\approx 1150-1700$\,\AA\ (dataset LEZCI5010, PI: Reindl). The spectra were retrieved from the MAST archive.

\begin{figure}
\centering
\includegraphics[width=\columnwidth]{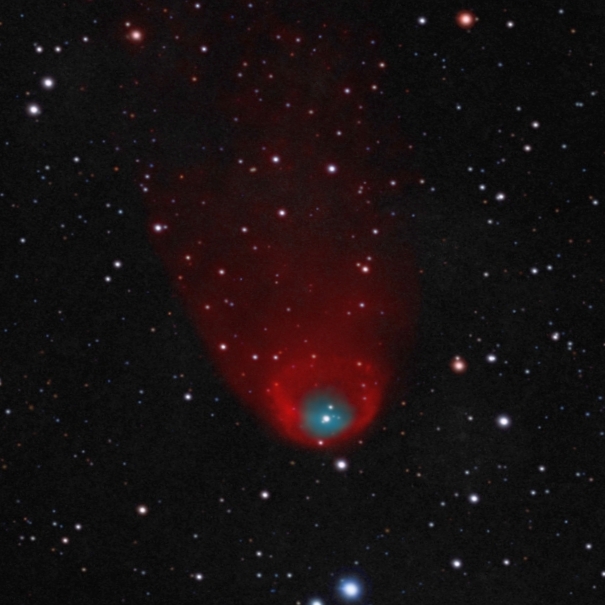}
\caption{RGB–H$\alpha$–[\ion{O}{iii}] composite image of K~1-6. North is up, east is to the left, and the field of view is approximately 17\arcmin{} $\times$ 17\arcmin{}.}
\label{fig:nebula}
\end{figure}

\subsection{Imaging}
Imaging of K~1-6 was conducted in April and May 2022 using twin APM TMB LZOS 152 mm refractors, each equipped with QSI6120 CCD camera and Astrodon filters. The observations were performed remotely at the e-EyE observatory in Fregenal de la Sierra, Spain. The dataset comprises broadband exposures in blue (12 $\times$ 300 s), green (12 $\times$ 300 s), and red (10 × 300 s), along with deep narrowband imaging in H$\alpha$ (132 $\times$ 900 s) and [\ion{O}{iii}] (64 $\times$ 900 s), resulting in a total integration time of 56 hours and 45 minutes.

\section{Results and discussion}
\subsection{The nebula and its morphology}
A faint nebulosity around K~1-6, nicknamed the Sleepy Eye Nebula by amateur astronomers, is visible in some survey images, but prior to this work, the deepest observations were those presented by \citet{frew11}, obtained with the 2-m Faulkes Telescope North. Their imaging showed that the nebula is asymmetric, with a brighter southern rim in H$\alpha$ + [\ion{N}{ii}], an off-centre central star, and a roughly circular morphology in [\ion{O}{iii}], more concentrated around the central star.

Although our images were obtained with smaller telescopes, the much longer total exposure times, 33 hours in H$\alpha$ and 16 hours in [\ion{O}{iii}], compared to 50 and 90 minutes in \citet{frew11}, allow us to reveal fainter structures. Our composite RGB H$\alpha$-[\ion{O}{iii}] image is shown in Fig. \ref{fig:nebula}, and represents the deepest image of this nebula obtained to date. The H$\alpha$ and [\ion{O}{iii}] images are displayed separately in Fig. \ref{fig:mosaic}.

In our data, it is seen much more clearly than in \citet{frew11} that the [\ion{O}{iii}] emission region is nearly circular ($2.05\arcmin \times 2.20\arcmin$), while the H$\alpha$ nebula is more asymmetric and extended ($3.18\arcmin \times 2.67\arcmin$). This ionisation structure is characteristic of a photoionised Strömgren sphere where the higher ionisation [\ion{O}{iii}] emission is confined closer to the central WD, and the lower ionisation H$\alpha$ emission extends further out. A previously unreported feature is also detected: a broad, conical tail extending more than 9\arcmin{} almost directly north of the central star. The tail appears to widen with increasing distance, consistent with a downstream wake. This overall morphology (combined with the fact that the axis of the tail is aligned nearly opposite to the proper motion of the system; see below) is highly suggestive of interaction with the surrounding interstellar medium (ISM) due to the high space velocity of the ionising source.

Assuming a distance to the object of $260 \pm 3$ pc (see below), this would translate to physical sizes of about 0.155 $\times$ 0.166 pc of the [\ion{O}{iii}] region, 0.241 $\times$ 0.202 pc of the brightest H$\alpha$ region, and the tail extending to more than $\sim$0.7 pc from the central star.

The nebula is also clearly detected in WISE mid-IR imaging \citep{2010AJ....140.1868W}. The dominant southern rim is visible in the W2 ($\approx$4.6 $\mu$m) and W3 ($\approx$12 $\mu$m) bands and extends even beyond the corresponding H$\alpha$ emission (Fig.~\ref{fig:mosaic}). We interpret this as the ionisation front moving forward through the undisturbed ISM. The detection of extended mid-IR emission indeed suggests the presence of heated interstellar dust. This ionised ISM has a higher pressure than its surroundings and thus expands at the sound speed, while the ionisation front itself moves with the speed of the star. Behind the star, the gas at first remains ionised by the star, but then recombines once beyond the Str\"omgren radius.  As H$^+$ recombines roughly four times slower than O$^{++}$, it extends further downstream.

\subsection{The white dwarf and the planetary nebula nature} \label{sec:white_dwarf}
To understand the source of ionisation and the nature of the nebula, we examine the UV spectrum from \textit{HST} and the optical nebular spectrum obtained with GTC. The UV counterpart of K~1-6 is clearly detected in both the \textit{GALEX} FUV and NUV images (Fig. \ref{fig:mosaic}). It lies close to the centre of the [\ion{O}{iii}] emission, and no other UV-bright source is present in the vicinity. This strongly indicates that the detected UV source is the ionising star responsible for the excitation of the nebula.

\begin{figure}
\centering
\includegraphics[width=\columnwidth]{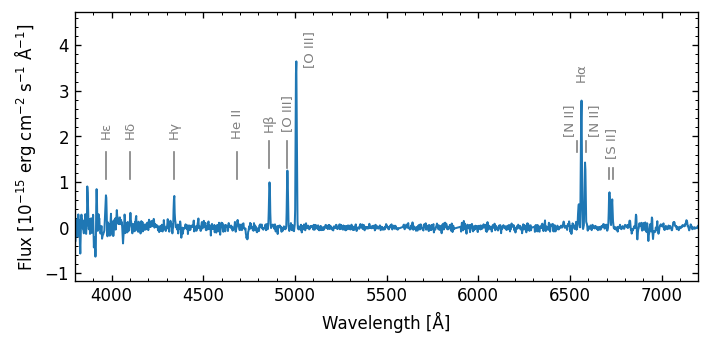}
\caption{GTC/OSIRIS spectrum of the nebula surrounding K~1-6 \citep{2023MNRAS.520..773R}. Prominent emission features are marked in grey.}
\label{fig:nebular_spectrum}
\end{figure}

\begin{figure}
\centering
\includegraphics[width=\columnwidth]{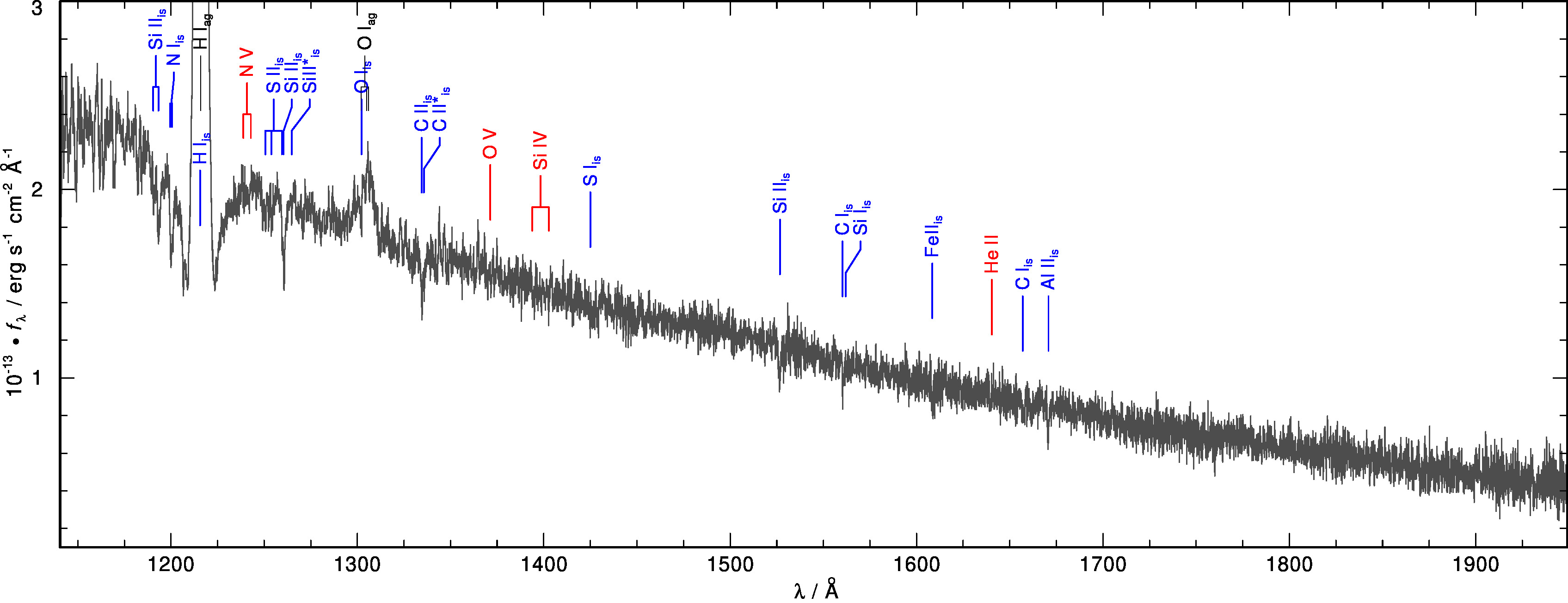}
\caption{HST/COS G140L spectroscopy of the WD. The locations of typically strong photospheric lines are highlighted in red, interstellar and airglow lines are highlighted in blue and black, respectively.}
\label{fig:hst2}
\end{figure}

\begin{figure}
\centering
\includegraphics[width=\columnwidth]{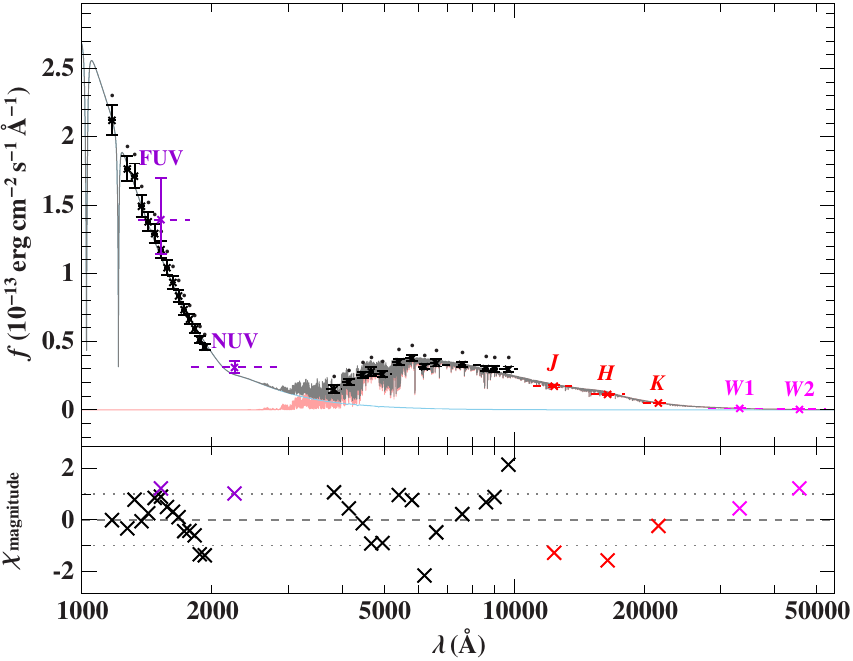}
\caption{Two-component spectral energy distribution fit. Top panel: Filter-averaged fluxes converted from observed magnitudes are shown in different colours (\textit{GALEX}: purple, 2MASS: red, WISE: magenta). The flux-averaged HST and \textit{Gaia} XP spectra are shown in black. The light blue and red lines correspond to the flux contribution of the WD and the cool star, respectively, and the grey line is the combined best fit to the observations. The model fluxes are degraded to a spectral resolution of $6\,\AA$. Bottom panel: Uncertainty-scaled difference between synthetic and observed magnitudes.}
\label{fig:sed}
\end{figure}

A lower limit on the temperature of the ionising source can be inferred from the nebular spectrum obtained with GTC (Fig.~\ref{fig:nebular_spectrum}). The spectrum shows strong Balmer emission together with [\ion{O}{iii}], [\ion{N}{ii}], and [\ion{S}{ii}], and also a faint but detectable \ion{He}{ii} $\lambda4686$ line (3-4 $\sigma$ detection). The presence of \ion{He}{ii} emission requires photons with ionising energy above 54.4~eV, implying an ionising star with an effective temperature, $T_{\rm eff} \gtrsim 54$~kK \citep[e.g.,][]{1994A&A...282..586M}.

On the other hand, no lines of He or photospheric metals can be detected in the HST spectra of the central star of K~1-6. This can be seen in Fig.~\ref{fig:hst2} and Fig.~\ref{fig:hst1}, where we highlight the locations of typically strong photospheric lines in red in addition to interstellar (blue) and airglow (black) lines. The absence of \ion{He}{ii} $\lambda\,1640$\,\AA\ suggests that the hot central star has an H-dominated atmosphere, and the lack of photospheric metals suggests a \Teff below 60\,kK \citep{Filiz+2024}. Thus, both nebular diagnostics and UV spectroscopy point to a relatively cool WD (compared to typical CSPN), with $T_{\rm eff}$ in the range of $\sim$54--60~kK.

\citet{frew11} already proposed that the central star of K~1-6 has a cool companion, based on the apparent superposition of a \textit{GALEX} UV source with a much redder star (dominating the optical and infrared photometry) and the pre-\textit{Gaia} astrometry. The separation between the catalogued \textit{Gaia} DR3 position of the optically dominant star \citep[][]{2016A&A...595A...1G,2023A&A...674A...1G} and the \textit{GALEX} UV source \citep{2017ApJS..230...24B} is 1.53\arcsec \citep[the astrometric precision of \textit{GALEX} is $\sim$0.5\arcsec; ][]{2007ApJS..173..682M}. In such a sparse field, the probability that two unrelated stars would lie this close together is very small. To quantify this, we estimated the local stellar density from \textit{Gaia} by counting stars brighter than a given magnitude within a radius of 30\arcmin{} around the object, and then calculated the expected number within the separation radius. The probability of finding by chance a star of $G=12.3$~mag, the brightness of the optically dominant star (or brighter) within 1.53\arcsec{} of the UV source is only 0.0054\%. Even considering all stars brighter than $G=20$~mag, the probability remains below 0.5\%. The association between the nebula and the bright optical star is further supported by the nebular morphology. The conical shape we discovered, indicative of interaction with the interstellar medium, aligns with the \textit{Gaia} proper motion of the optical star (the proper motion vector corrected for Galactic rotation is within 10\textdegree{} of the uncorrected one; observed PA = 180.41$\pm$0.08\textdegree{}, $\mu$ = 28.19$\pm$0.05 mas\,yr$^{-1}$, corrected PA = 190.30$\pm$0.10\textdegree{}, $\mu$ = 32.90$\pm$0.08 mas\,yr$^{-1}$), which points nearly directly opposite the direction of the nebular cone (see Sect. \ref{sec:multiplicity}). This strongly supports a physical connection between the nebula, the UV source, and the optically dominant star, and allows us to adopt the \textit{Gaia} DR3 distance of the optically dominant component to infer the intrinsic properties of the UV source.

Given the \textit{Gaia} distance to the system (adopted 260 pc in this work; Sect. \ref{sec:multiplicity}), the UV fluxes place strong constraints on the luminosity of the hot component. We performed a two-component fit to the SED (see Fig.~\ref{fig:sed}), employing the $\chi ^2$ fitting routine described in \cite{Heber+2018} and \cite{Irrgang+2021}, the \textit{GALEX} FUV and NUV magnitudes together with the \textit{HST} spectrum (flux-averaged in 50\,\AA{} bins), \textit{Gaia} low-resolution spectra and infra-red photometry. For the cool star, we used PHOENIX models calculated by \cite{Husser+2013}, and for the hot central star, the pure H model grid computed by \cite{Reindl+2023}. We fixed the $T_{\rm eff}$ of the hot star to 57\,kK, and then let the angular diameter, reddening, and surface ratio of both stars, as well as the $T_{\rm eff}$ of the cool companion, vary freely. By that we derive a reddening of $E(44-55)=0.09$\,mag\footnote{\cite{Fitzpatrick+2019} employs $E(44-55)$, which is the monochromatic equivalent of usual $E(B-V)$, using the wavelengths 4400\,\AA\ and 5500\,\AA, respectively. For high effective temperatures such as for the stars in our sample $E(44-55)$ is identical to $E(B-V)$.}. From the angular diameter, $\Theta$, and the zero point corrected \textit{Gaia} parallax, $\varpi$, the radius, $R$,  of the hot central star can be calculated via $R = \Theta/(2\varpi)$, and the luminosity, $L$, is obtained via $L/L_\odot = (R/R_\odot)^2(T_\mathrm{eff}/T_{\mathrm{eff},\odot})^4$. We find $R=0.0177$\,\Rsol\ and $L=3.0$\,\Lsol, indicating that the hot star is indeed a WD. Using the effective temperature and luminosity of the WD, we interpolated its mass using the evolutionary tracks from \citet{Renedo+2010}, and find $M_{\mathrm{WD}}=0.58$\,\Msol. Considering effective temperatures between $54-60$\,kK, this corresponds to a cooling age of $1-2$\,Myrs. 

At such an advanced evolutionary stage, any material ejected during the AGB or early post-AGB phase would have long since expanded, diluted, and merged into the ambient ISM. For a typical nebular expansion velocity of 20--30 km\,s$^{-1}$, material travels a parsec within $\sim10^5$ yr, and its surface brightness drops below detectability on timescales of only $\sim(2-5)\times10^4$ yr. This is orders of magnitude shorter than the inferred post-AGB age of the WD in K 1-6. Such a large mismatch in timescales makes it implausible that any coherent PN shell originating from the progenitor could survive to the present epoch, even in highly favourable cases.

There is, however, observational and theoretical evidence that PNe can be partially rebrightened through interaction with the interstellar medium \citep[e.g.,][]{1990ApJ...360..173B,2007MNRAS.382.1233W,2010PASA...27..220W,2016MNRAS.457....9C}, particularly when the central star moves supersonically through the local ISM. Objects such as Sh 2-188 illustrate the early stages of this process, showing a compressed, filamentary arc facing the direction of motion, a partially preserved shell in the opposite direction, and a broad downstream tail \citep{2006MNRAS.366..387W}. Similarly, HFG 1 shows a well-defined bow-shock structure and a long cometary wake \citep{2009MNRAS.396.1186B}. In all of these systems, however, the central stars are still relatively young post-AGB objects, with luminosities of several hundred solar and evolutionary ages below a few $\times10^5$ yr, and their nebulae remain dynamically connected to a recognisable PN shell.

Although morphologically K 1-6 does appear to some extent like these interacting PNe, it differs from them in other parameters. Its WD has a luminosity of only $\approx 3.0\,L_{\odot}$, consistent with a post-AGB age >$10^6\,\text{yr}$, placing it far beyond the regime where PN material can remain structurally or dynamically intact. Moreover, the radial velocity of the nebular emission lines provides a crucial independent diagnostic of the gas origin.

The nebular heliocentric radial velocity of $-6\,\text{km s}^{-1}$ reported by \citet{2023MNRAS.520..773R}\footnote{The authors stated typical velocity error of $\pm$ $15\,\text{km s}^{-1}$.} is far from the stellar systemic velocity of $\sim$${-51.6\,\text{km s}^{-1}}$ (obtained from our NOT/FIES measurements; consistent with \textit{Gaia} DR3 value of ${-55\,\text{km s}^{-1}}$). For an expanding PN shell, the bulk nebular material would share the stellar velocity, resulting in line centroids significantly shifted relative to the ISM. Instead, the observed nebular velocity is inconsistent with an expanding PN shell but is consistent with the quiescent local ISM. This disparity strongly suggests that the emitting gas is predominantly interstellar in origin, swept up by the motion of the K 1-6 system, rather than remnant ejecta.

Adopting the $\textit{Gaia}$ DR3 proper motion and radial velocity of the optically dominant star, which is physically associated with the WD and nebula, the three-dimensional space velocity relative to the local ISM can be determined after correcting for Galactic rotation. Using the routines of \citet{2023A&A...680A..99M}, we find that the system moves at approximately ${60\,\text{km s}^{-1}}$ with respect to the local Galactic velocity field (i.e., the expected bulk velocity of the surrounding interstellar medium). Such a velocity corresponds to a Mach number of $\approx5$ in warm neutral/warm ionised ISM.  Ultimately, the observed properties indicate that K 1-6 is not a surviving planetary nebula, but a photoionised Strömgren sphere in the local ISM.

\begin{figure}
\centering
\includegraphics[width=\columnwidth]{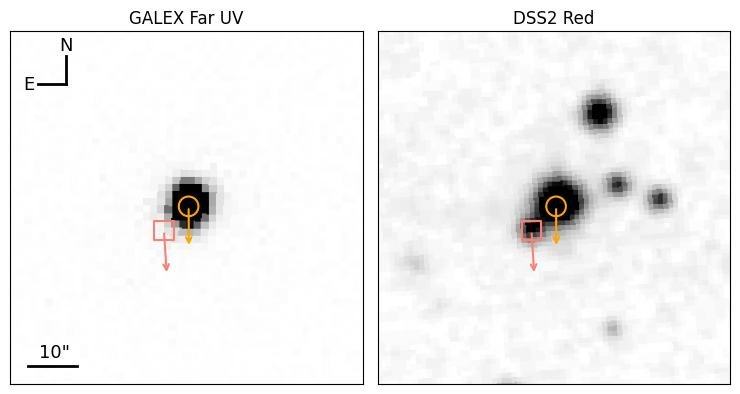}
\caption{The 72\arcsec{} field around K~1-6 (indicated by the orange circle). The position of the common proper motion companion, located at a similar distance, is marked by the pink square (see text for details). Proper motion vectors for both objects are shown as arrows. The left panel displays the \textit{GALEX} Far-UV image, while the right panel shows the DSS2 Red image.
}
\label{fig:proper_motions}
\end{figure}

\subsection{Multiplicity of K~1-6}\label{sec:multiplicity}

The binary nature of the central star of K~1-6 and its relation to the nebula was firmly established in the previous sections. In addition, \textit{Gaia} DR3 data reveal a common proper motion companion to the central binary source (Gaia DR3 2288467190738512256; $G = 17.0$ mag; Fig. \ref{fig:proper_motions}), as already noted by \citet{2023Ap&SS.368...34A}, located at an angular separation of 7.07\arcsec{}. The two \textit{Gaia} sources exhibit similar proper motions: in right ascension, $-0.195 \pm 0.037$ and $-1.955 \pm 0.100$ mas\,yr$^{-1}$, and in declination, $-28.186 \pm 0.045$ and $-30.160 \pm 0.160$ mas\,yr$^{-1}$, for the central star of K~1-6 and the companion, respectively. Their parallaxes are also consistent, measured as $3.8637 \pm 0.0302$ and $3.8187 \pm 0.0894$ mas (well within 1 $\sigma$), indicating that both objects lie at similar distances.

Assuming a distance of $260 \pm 3$ pc (derived as the average from the inverse of their parallaxes\footnote{{The elevated RUWE of K 1-6 suggests that the formal \textit{Gaia} parallax uncertainty may be underestimated \citep{2025OJAp....8E..62E}. According to equation~3 of \citet{2025OJAp....8E..62E}, the corresponding uncertainty inflation factor is approximately 3.4.}}), the projected physical separation between the central star of K~1-6 and its companion is $1841 \pm 23$ au. This separation corresponds to an orbital period on the order of tens of thousands of years. The presence of this wide companion implies that the system is a hierarchical triple.

We note that the central star of K~1-6 is not included in the catalogue of wide binaries by \citet{2021MNRAS.506.2269E}. Although the projected separation and parallaxes satisfy their selection criteria (see their Section 2), the proper motions do not fully meet the requirement of being consistent with a Keplerian orbit. This discrepancy is expected if the system is not a simple wide binary but instead consists of a closer inner binary accompanied by a wider third component. In such a configuration, the orbital motion within the inner pair perturbs the proper motion of the central source, causing it to fail the wide-binary consistency criteria. 

\textit{Gaia} DR3 itself lists the central star of K~1-6 in the non-single-star catalogue with an astrometric acceleration solution \citep[][]{2023A&A...674A...9H}. This solution suggests an orbital period on the order of several hundred to several thousand days. The Renormalised Unit Weight Error (RUWE) is also elevated, with a value of 2.93, well above the typical single-star threshold of $\sim$1.25 \citep{2022MNRAS.513.2437P}. This astrometric signal traces the orbit of the cool component and the WD, also supporting the idea that the inner binary orbital motion is influencing the proper motion of the source. {This system will likely receive a fully resolved orbital solution in the upcoming \textit{Gaia} DR4.}

In relation to the third wide companion, we note that \citet{2025MNRAS.543.3035C} stated that it is unclear which of the two \textit{Gaia} common–proper-motion sources is associated with the UV detection. They stated that while \citet{frew11} favoured the brighter one, \citet{2021A&A...656A.110C} and \citet{2021A&A...656A..51G} pointed to the fainter source. This appears to be incorrect: both of the latter works explicitly identify the brighter \textit{Gaia} source, Gaia DR3 2288467186442571008, as the central star. \citet{2025MNRAS.543.3035C} further claimed that although the \textit{GALEX} source lies closer to the brighter component, its footprint encompasses both stars. This is also not supported by the data. Two \textit{GALEX} tiles cover the field, one comprising three NUV and three FUV visits between September 24 and November 12, 2005, and another with two visits per filter between September 10 and November 1, 2006. In both epochs, K~1-6 falls near the edge of the field, resulting in some distortion; nevertheless, especially in the FUV images, the association with the fainter companion can be confidently ruled out (see Fig.~\ref{fig:proper_motions}). The association with the brighter counterpart is also further confirmed by the coordinates of the UV source observed by HST.

The characterisation of the third wide companion in K~1-6 is complicated by a background star (Gaia DR3 2288467190737047424), detected by \textit{Gaia} at a separation of 0.96\arcsec{}. It is slightly brighter than the third component of the K~1-6 system itself ($G = 16.7$ mag, $BP = 16.9$ mag, $RP = 15.3$ mag). However, this object has a much smaller parallax ($0.16 \pm 0.17$ mas) and a proper motion of only 1.89 mas\,yr$^{-1}$, indicating that it is a distant background source and not physically associated with K~1-6. Nonetheless, its proximity may affect both current and future observations of the wide companion. For example, photometric measurements such as those from 2MASS include blended light from both sources, and should therefore be interpreted with caution. As a result, we cannot reliably determine the properties of the third K~1-6 component (e.g., via SED fitting; even the $BP-RP$ colour is unavailable). Assuming solely its $G$ magnitude and a distance of 260 pc, the calculated absolute magnitude suggests it is an M2--3 dwarf, following \citet{2013ApJS..208....9P}.

\subsection{Stellar parameters of the optically dominant component}\label{sec:cool}

Having established that the ultraviolet source is spatially coincident (and physically related) with a redder, \textit{Gaia}-detected star, we next characterise the properties of this optically dominant component of the central binary. From the SED fit (Sect. \ref{sec:white_dwarf} and Fig. \ref{fig:sed}), for the cool star, we find \Teff$=4590$\,K, $R=1.50$\,\Rsol\ and $L=0.9$\,\Lsol. 

Our high-resolution spectra, in particular the NOT/FIES data from May 4, 2023, were analysed using the open-source iSpec framework for stellar parameter determination \citep{iSpec, Blanco-Cuaresma19}. The tree spectra obtained that night were corrected for their radial-velocity shifts and co-added to increase the signal-to-noise ratio. The stellar parameters were derived through synthetic spectral fitting, using a $\chi^2$ minimisation between the observed and synthetic spectra in the wavelength range 4\,800--6\,200~\AA{}. For the synthesis, we employed the radiative transfer code Turbospectrum \citep{Turbospectrum98,Turbospectrum12}, MARCS model atmospheres \citep{Gustafsson2008}, the solar abundances of \citet{Asplund09}, and version~6 of the Gaia-ESO spectral line list \citep{Heiter2021}. We assumed in this analysis that just one star contributes to the observed spectra.

The best-fitting parameters are:
$T_{\rm eff} = 5000 \pm 100$~K;
$\log g = 4.2 \pm 0.2$;
[Fe/H] $= -0.25 \pm 0.10$;
[$\alpha$/Fe] $= 0.2 \pm 0.10$;
$v_{\rm mic} = 3.0 \pm 0.4$~km\,s$^{-1}$;
$v_{\rm mac} = 3.4$~km\,s$^{-1}$ (fixed from empirical relations);
$v \sin i = 16.0 \pm 0.6$~km\,s$^{-1}$. The derived values are roughly consistent with those expected for a $\sim$K2V star\footnote{\url{https://www.pas.rochester.edu/~emamajek/spt/K2V.txt}} \citep{2013ApJS..208....9P}. It is worth noting that the microturbulence value is significantly larger than typically measured in K dwarfs ($\approx$ 1 km s$^{-1}$). This provides a first indication that the stellar parameters derived here are affected by important systematic uncertainties, and that the internal errors quoted are likely underestimated.

Another indication of the difficulty in obtaining accurate stellar parameters is the discrepancy between the effective temperature derived here and that obtained from the SED fit. However, the analysed spectrum was acquired near the maximum brightness of the star (see Sect. \ref{sec:variability}), and even a simple estimate based on V–K colours between maximum and minimum brightness yields a wide range of effective temperatures, from 4430 to 5320 K (G9V–K5V; \citealt{2013ApJS..208....9P}). Using B–V colours suggests even lower temperatures (4200–4330 K; $\sim$K6V), although the bluest colours may be affected by the WD companion.

\begin{figure}
\centering
\includegraphics[width=\columnwidth]{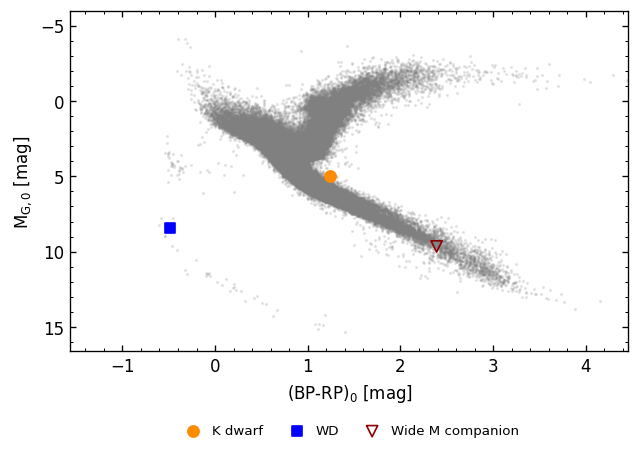}
\caption{Position of the components of K~1-6 in the \textit{Gaia} colour--magnitude diagram. The optically dominant component is marked by an orange circle, the WD by a blue square, and the wide companion by an open red triangle (its colour is inferred from its assumed spectral type). The background shows a well-characterised sample of \textit{Kepler} stars from \citet{2025A&A...696A.243G}.
}
\label{fig:gaia_cmd}
\end{figure}

The inferred parameters of the cool component broadly agree with the \textit{Gaia} photometry and astrometry: its position in the colour–magnitude diagram is consistent with a mid-K dwarf (Fig. \ref{fig:gaia_cmd}), although the star appears slightly overluminous compared to a normal single main-sequence star. This overluminosity may support the relatively low $\log g$ derived here. Indeed, similar overluminosity has been reported in other systems hosting fast-rotating K dwarfs in late planetary nebulae \citep{2026arXiv260323756B}.

Using these tentative stellar parameters, and exploring their full range, we attempted to derive key elemental abundances {that may carry signatures of prior mass transfer from the progenitor of the WD, which evolved through the AGB phase before becoming the present-day companion}. {Material accreted by the K dwarf during this phase could have altered its surface composition and increased its angular momentum. We therefore searched for abundance patterns associated with AGB nucleosynthesis.}

\begin{figure}
\centering
\includegraphics[width=\columnwidth]{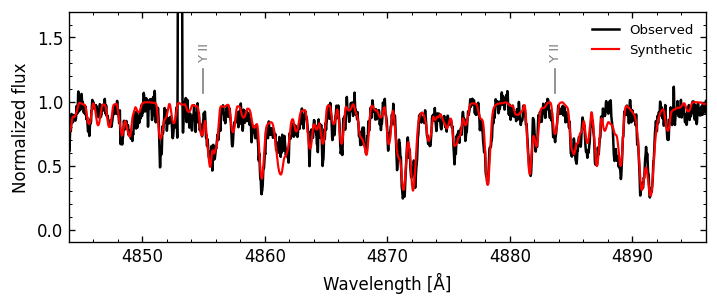}
\includegraphics[width=\columnwidth]{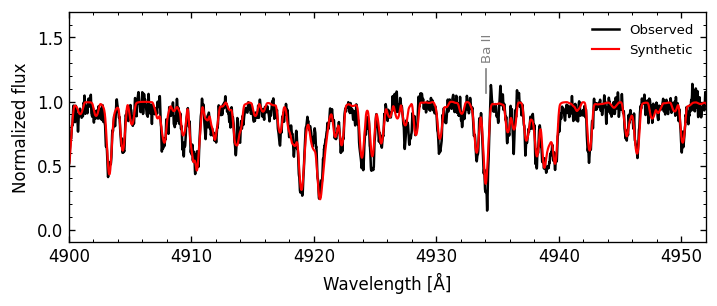}
\includegraphics[width=\columnwidth]{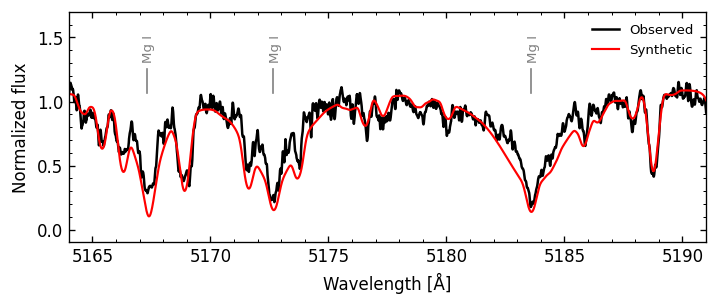}
\includegraphics[width=\columnwidth]{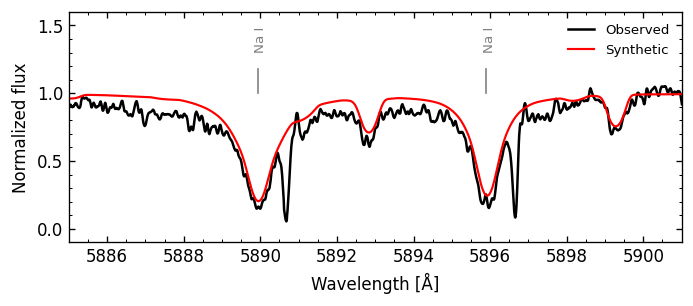}
\caption{Example fit of the synthetic spectrum (red; see text) to the observed spectrum (black; co-added NOT/FIES data from May 4, 2023). The Ba, Y, Mg, and Na lines discussed in the text are marked.
}
\label{fig:parameters}
\end{figure}

We first examined C and N by fitting CH, C$\rm _2$, and CN molecular features in the NOT/FIES spectra. The CH features indicate a low carbon abundance, while C$\rm _2$ bands also suggest carbon depletion, and the CN features do not provide evidence for nitrogen enhancement, as would be expected if CN-cycle processing had occurred. We then analysed Mg, particularly using the Mg triplet lines (Fig. \ref{fig:parameters}). Their wings suggest either that the adopted surface gravity is still too high or that Mg is underabundant.  
Mg depletion can occur in massive AGB stars, but it is typically accompanied by strong Al production. However, no Al enhancement is detected in the spectrum.

The Na abundance, derived from the Na D lines and other features, appears to be consistently enhanced (Fig. \ref{fig:parameters}). In AGB stars, this may be accompanied by O depletion. The oxygen abundance could only be derived from the O triplet, since the forbidden [\ion{O}{i}] line at 6330 \AA{} is affected by telluric contamination. However, the triplet does not indicate any clear oxygen depletion. Moreover, given the inconsistencies found for other elements, this result should be treated with caution. Figure \ref{fig:parameters} also shows spectral fits in two representative wavelength windows, including the Ba II 4934.1 \AA{} and Y II 4883.7 \AA{} lines. These $s$-process features are reasonably well reproduced without invoking any enrichment, contrary to previous suggestions in the literature \citep{2013MNRAS.436.2082T}. However, these elements are detected through ionised lines, which are highly sensitive to uncertainties in stellar parameters and to non-standard atmospheric and radiative transfer effects. Consequently, we consider these abundance indicators unreliable.

In summary, despite exploring a wide range of stellar parameters, we are unable to derive a fully consistent abundance pattern for this star. This likely reflects the peculiar nature of its atmosphere, possibly linked to its extreme stellar activity.

\subsection{Activity of the cool component}\label{sec:activity}

\begin{figure}
\centering
\includegraphics[width=\columnwidth]{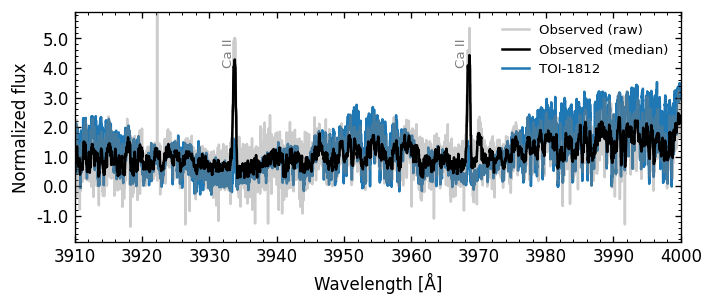}
\includegraphics[width=\columnwidth]{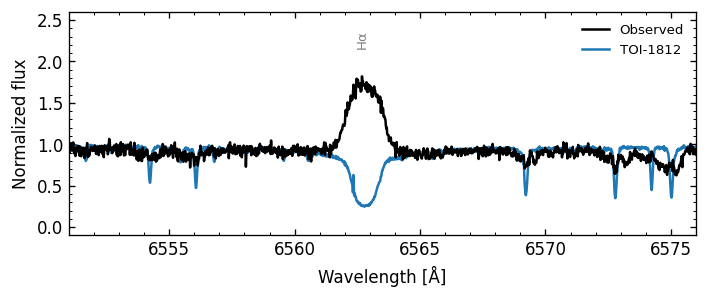}
\includegraphics[width=\columnwidth]{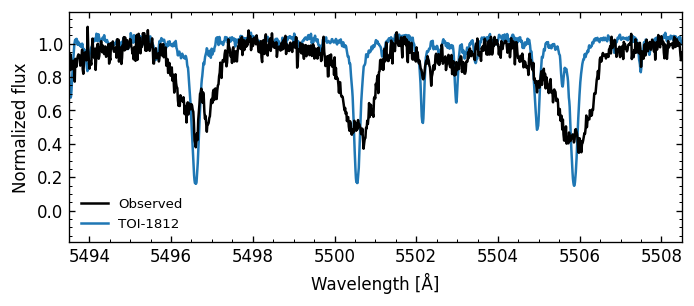}
\includegraphics[width=\columnwidth]{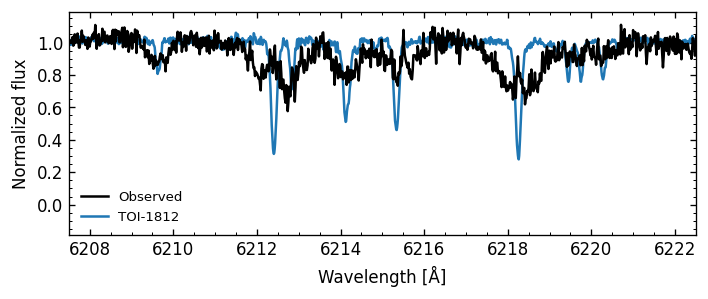}
\caption{Emission lines in the TNG/HARPS-N spectrum from June 17, 2025 (upper two panels) and emission cores of Fe I lines (lower two panels), compared with the spectrum of TOI-1812, a K dwarf with similar parameters and from the same instrument. A running median filter has been applied to the Ca II line regions.
}
\label{fig:emission_lines}
\end{figure}

The spectroscopic parameters should be interpreted with caution, as the data clearly indicate that the cool companion is chromospherically active. Figure~\ref{fig:emission_lines} shows strong H$\alpha$ emission together with the \ion{Ca}{ii} doublet, both of which are classical indicators of magnetic activity. We measured the radial velocities of H$\alpha$ and the \ion{Ca}{ii} lines in the HARPS-N spectra from June and September 2025. Within uncertainties of $\sim$2.5~km\,s$^{-1}$, these are consistent with the radial velocity of the optically dominant component. The same behaviour is seen in the NOT/FIES spectra.

Fig. \ref{fig:emission_lines} shows remarkable emission in the cores of several \ion{Fe}{i} lines as well. Although such features are only sparsely documented in the literature (except for T Tauri stars), \citet{Vieytes2025} demonstrated through tailored models including non-local thermodynamical equilibrium transfer consistently with an extended chromosphere, that some \ion{Fe}{i} lines are indeed expected to appear in emission in the spectra of active G dwarfs. They also emphasise that, in addition to the H Balmer lines and \ion{Fe}{i} lines, the \ion{Mg}{i} triplet and \ion{Na}{i} D line profiles are significantly affected by stellar activity. Moreover, \citet{Vieytes2009} pointed out that the chromospheric thermal structure changes drastically with activity, making it essential to rely on quasi-simultaneous observations when constructing atmospheric models for active stars. This likely explains the difficulty in modelling the spectrum with standard models discussed in the previous section and raises doubts about the reliability of the derived abundances.

In addition to these activity indicators, we note that \citet{2024A&A...689A.103Z} and \citet{2025A&A...694A.161S} reported the detection of multiple flares from the central star of K~1-6 in the TESS 20\,s cadence data. The presence of such flares provides further independent evidence for an efficient magnetic dynamo operating in the cool component.

\citet{frew11} reported X-ray emission from the system detected with \textit{ROSAT}. Re-scaling to a distance of 260 pc and adopting $E(B-V)=0.1$ mag, we obtain and X-ray luminosity of $L_X \sim 1.81 \times 10^{30}$ erg\,s$^{-1}$. This places the source at the upper end of X-ray luminosities observed in K-type stars, typically $10^{26-29}$ erg\,s$^{-1}$, though rapidly rotating or young stars can reach $\sim10^{30}$ erg\,s$^{-1}$.

Given the estimated system age of several Gyr (cooling age of the WD, see Sect. \ref{sec:white_dwarf}, and the progenitor age\footnote{{The initial mass of the current WD progenitor is estimated to be $\sim$1.6$\pm$0.3\,M$_\odot$ under the single-star approximation \citep[e.g.,][]{2018ApJ...860L..17E,2024MNRAS.527.3602C}. This corresponds to a pre-WD lifetime of approximately 3.5--4\,Gyr, although the exact value depends on the adopted stellar evolution models and parameters such as metallicity.}}), such a high X-ray luminosity would be difficult to reconcile with a normal, slowly rotating single K dwarf. However, many parameters indicate the cool component of K 1-6 is highly peculiar. Alternatively, part or all of the X-ray emission may originate from the hot WD, which is expected to emit soft X-rays due to its hot photosphere \citep[e.g.,][]{1993MNRAS.264...16B,1996A&A...316..147F}. The observed luminosity is consistent with this possibility, although the current data do not allow us to disentangle the contributions of the two components.

\subsection{Photometric and radial-velocity variability}\label{sec:variability}

\begin{figure}
\centering
\includegraphics[width=\columnwidth]{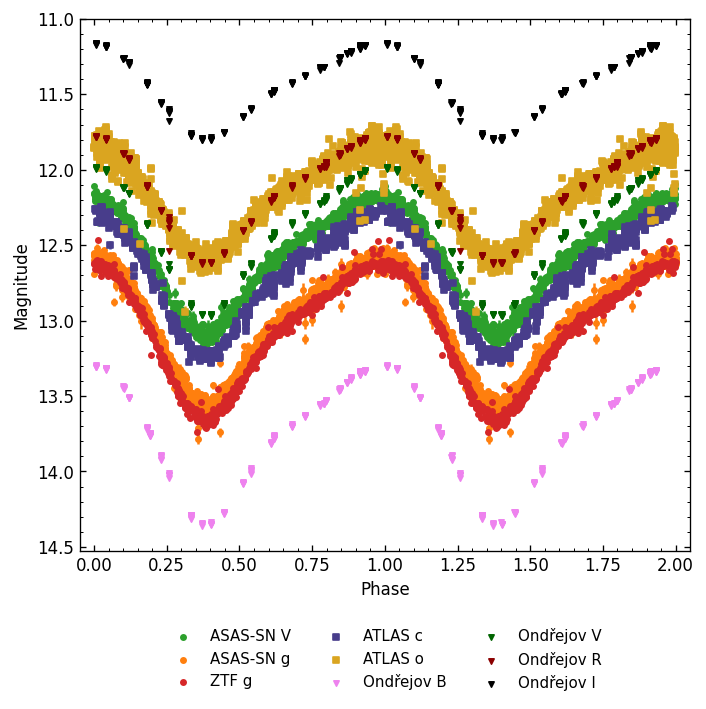}
\includegraphics[width=\columnwidth]{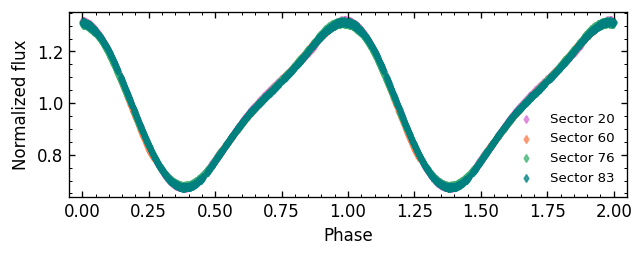}
\caption{Phased light curves of K~1-6. \textbf{Upper panel:} Ground-based observations in various photometric bands, folded with a 21.305~d period. \textbf{Lower panel:} TESS photometry from selected sectors in Cycles 2, 4, and 5.}
\label{fig:photometry}
\end{figure}

Analysis of the available photometric data for the central star of K~1-6 reveals periodic variability with a period of $P$ = 21.305~days. The amplitude increases toward shorter wavelengths, reaching a peak-to-peak amplitude of approximately 1~mag in the $B$ band and about 0.5~mag in the $I$ band. The phased light curves, shown in Fig.~\ref{fig:photometry}, exhibit an asymmetric, sawtooth-like shape with a steeper decline branch. This waveform can be well reproduced using a Fourier series composed of three sine components with periods of $P$, $P/2$, and $P/3$. 

The shape of the variability appears stable over the $\sim$13-year time span covered by the photometric data analysed here. This stability is evident even at the higher precision of the TESS observations, as illustrated by the selected sectors (over Cycles 2, 4, and 5; i.e., over > 5 years) shown in the lower panel of Fig.~\ref{fig:photometry}, and likely persists over an even longer timescale, as suggested by comparison with the older Northern Sky Variability Survey (NSVS) data presented by \citet{frew11}, obtained between April 1999 and March 2000. We note that no other significant periodicities, either shorter or longer than 21.3 days, were detected in the analysed photometric data {(see Fig. \ref{fig:periodograms})}.

\begin{figure}
\centering
\includegraphics[width=\columnwidth]{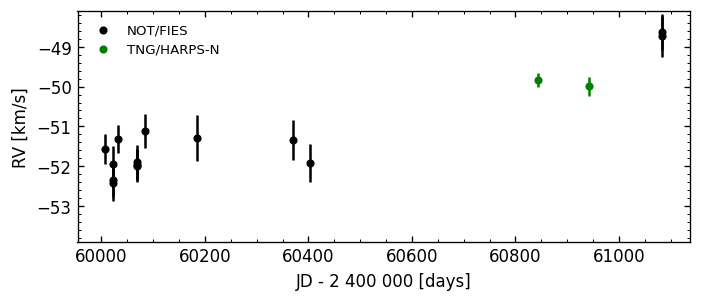}
\includegraphics[width=\columnwidth]{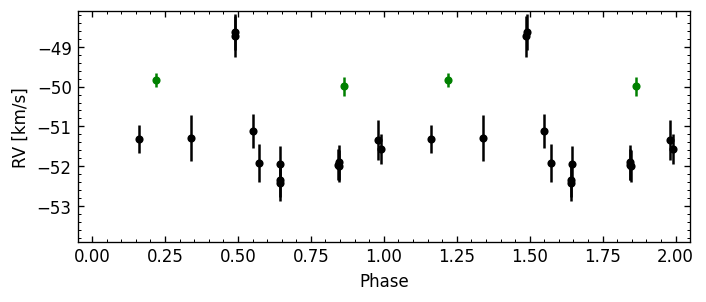}
\caption{Radial velocities of K~1-6. \textbf{Upper panel:} Radial velocities measured from NOT/FIES and TNG/HARPS-N spectra. \textbf{Lower panel:} Radial velocities folded with the photometric period of 21.305~d.}
\label{fig:rvs}
\end{figure}

Radial velocity measurements of the cool component of K~1-6, derived from FIES and HARPS-N spectroscopy and phased with the same 21.305-day period, are shown in Fig.~\ref{fig:rvs}. We used iSpec \citep{iSpec, Blanco-Cuaresma19} to derive radial velocities from each spectrum, fitting a Gaussian to the cross-correlation function (CCF) computed with a K5 spectral mask. While there is a marginal indication of variability with this period, the amplitude is low, around 1~km\,s$^{-1}$, and the radial velocities are dominated by the long-term trend (consistent with the possible thousands of days orbit of the K dwarf and WD; {in line with \textit{Gaia} acceleration solution}). This suggests that the observed photometric period is unlikely to correspond to orbital variability. For comparison, assuming a 0.82~M$_\odot$ K2-type dwarf (Sect. \ref{sec:cool}) and a 0.58~M$_\odot$ WD companion (Sect. \ref{sec:white_dwarf}) in a circular orbit with a 21.3-day period, one would expect a radial velocity amplitude of $\sim$36~km\,s$^{-1}$ for an edge-on system ($i=90^\circ$), and still $\sim$9~km\,s$^{-1}$ for a nearly face-on orientation ($i=15^\circ$). The photometric variability is therefore most likely intrinsic to the cool component.

\subsection{Origin of the variability and activity}\label{sec:rotation}
The non-detection of significant short-term orbital radial velocity variations in our observations suggests that the optically dominant component of K 1-6 is not part of any short-period binary. The observed photometric variability is likely rotational and not due to irradiation, since the distance between the hot component and the K2 dwarf is too large. Among planetary nebula central stars, several other wide binaries with chromospherically active companions are known \citep{2013MNRAS.436.2082T,miszalski13,jones22}. However, none show such strongly asymmetric photometric variability. 

\begin{figure}
\centering
\includegraphics[width=\columnwidth]{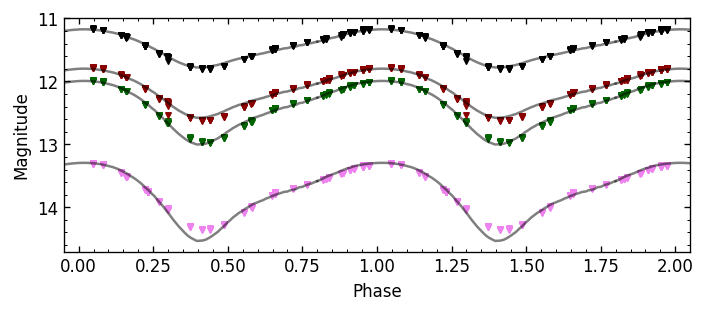}
\caption{Ondřejov $B$, $V$, $R$, and $I$ light curves of K~1-6 with the best-fit \textsc{phoebe2} model (gray lines). The colours are the same as in Fig. \ref{fig:photometry}.}
\label{fig:phoebe}
\end{figure}

To test whether the variability can be attributed to spots, we constructed a simple rotating single-star model with parameters consistent with a K2V star using \textsc{phoebe2} \citep{phoebe2,phoebe3,phoebe4,phoebe5}. Allowing for two spots on the surface of the star and fitting for their angular radii, longitude (their latitude was fixed to 90$^\circ$, maximising the amplitude of variability) and relative effective temperature (compared to that expected for no spots), we could reasonably reproduce both the light curve morphology and colour-dependence of the amplitude (Fig. \ref{fig:phoebe}). The spots required are large (radii of approximately 30$^\circ$ and 60$^\circ$) and high contrast (relative effective temperatures of approximately 0.7 and 0.8, respectively). The model also predicts radial-velocity variations with amplitudes of order 1~km\,s$^{-1}$, although strongly non-sinusoidal with phase. 

It is important to note that this single-star model does not account for any flux contamination or dilution from other component(s) in the system. As a result, the actual spots could be even more extreme. One would expect such dilution to be strongest in the bluer bands, which are, in fact, the bands where the model fit is poorest, with the predicted amplitudes of variability being larger than observed.

The relative temperature contrast is broadly consistent with expectations for a star of this spectral type, and the inferred spot coverage is not implausible, having been observed in other objects \citep[see, e.g.,][and references therein]{2005LRSP....2....8B}. However, this interpretation requires the spots to remain stable over a prolonged interval (at least 13--25 years), significantly longer than typically observed for normal K dwarfs \citep[see, e.g.,][]{2017MNRAS.472.1618G,2022ApJ...924...31B}. 

\begin{figure}
\centering
\includegraphics[width=\columnwidth]{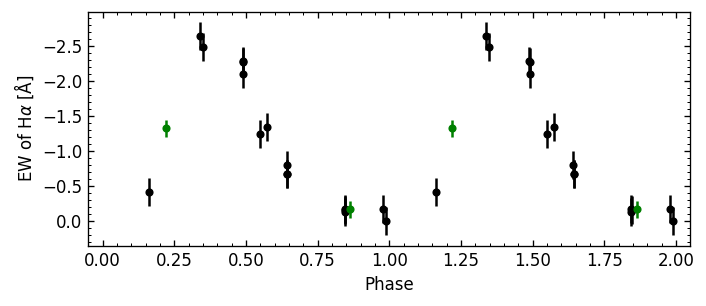}
\caption{Equivalent width of the H$\alpha$ line folded with the 21.3-day photometric period. The colours are as in Fig. \ref{fig:rvs}.
}
\label{fig:halpha}
\end{figure}

The equivalent width of H$\alpha$ (and also the emission in the core of the \ion{Ca}{ii} H\&K lines, although only marginally detected in the NOT data because of the low S/N) also follows the 21.3-day period (Fig.~\ref{fig:halpha}). This behaviour is consistent with rotational modulation of chromospheric activity, where active regions on the stellar surface contribute enhanced H$\alpha$ emission as they rotate across the visible hemisphere. In particular, the increase in equivalent width near photometric minimum suggests that the dominant active region is most visible at these phases, leading to stronger line emission.

The presence of strong chromospheric emission lines (Sect. \ref{sec:activity}) and high X-ray luminosity further supports the interpretation that the star is magnetically active. In this framework, several otherwise puzzling properties can be understood as consequences of activity. In particular, large spot coverage can affect inferred parameters and also distort spectral line profiles.

\begin{figure}
\centering
\includegraphics[width=\columnwidth]{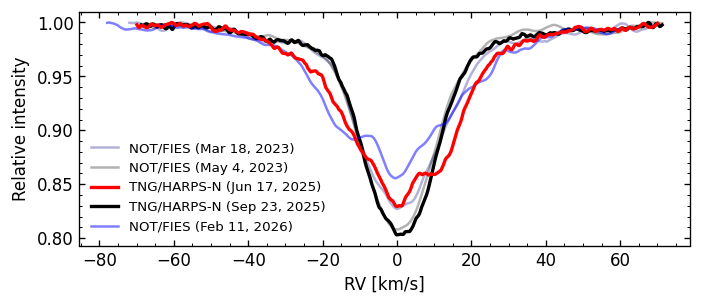}
\caption{Cross-correlation functions (CCFs) from the two co-added NOT/FIES observations and the two TNG/HARPS-N datasets. All CCFs are velocity-shifted so that the primary peak is centred at 0 km\,s$^{-1}$ for comparison.
}
\label{fig:ccf}
\end{figure}

Some of the CCF profiles indeed appear complex (Fig.~\ref{fig:ccf}), in some cases resembling those of a double-lined spectroscopic binary (SB2). This feature cannot be attributed to either the hot component of K~1-6, whose contribution in the optical is negligible, or to the wide companion, which lies outside the fiber aperture. 

A clear double-peaked CCF is seen in one HARPS-N epoch (June 2025; Fig.~\ref{fig:ccf}), while a second HARPS-N spectrum obtained 98 days later (September 2025) shows only a single, slightly narrower peak. Most FIES observations also show a single peak, although several of them were obtained under suboptimal conditions. The FIES spectrum from February 2026 again shows a double-peaked profile, but with the secondary component located on the opposite side of the main peak compared to the June 2025 HARPS-N spectrum. With the limited dataset available, we note that the double-peaked CCF tends to appear when the star is fainter, that is, when a significant fraction of its surface may be covered by spots if the rotational modulation scenario is correct, while it is not detected close to maximum brightness. The HARPS-N spectrum showing the double peak was obtained before the photometric minimum, whereas the FIES spectrum with a similar feature was taken after the minimum.

We suspect that the double-peaked CCF is caused by line-profile deformation due to large starspots on the surface of the K2 dwarf (likely combined with emission cores in some of the lines). Cool spots can distort the line profiles and produce asymmetric or even multi-peaked CCFs (see Appendix \ref{app:line_profiles}), which may mimic the signature of an SB2 system. This effect is expected to be strongest when the spotted regions cover a significant fraction of the visible hemisphere, i.e., close to photometric minimum, and weaker near maximum brightness. This behaviour is qualitatively consistent with our observations, where the complex CCF profiles appear preferentially at fainter phases of the light curve. In this scenario, the changes in the position of the secondary peak between epochs would not trace orbital motion (as one would expect for an SB2 binary), but rather the changing visibility of active regions as the star rotates. Higher cadence, high signal-to-noise spectroscopy covering the full photometric cycle could further verify whether the CCF variability follows the rotational period.

The alternative explanation as a genuine SB2 signature would imply that the system is quadruple, but it raises several difficulties. The radial velocity of the dominant component does not show significant variations, while the putative secondary peak shifts noticeably between epochs. This would require an unusual orbital configuration, and it is unclear what type of companion could produce a detectable CCF peak without inducing a comparable radial-velocity signal in the primary. The observed velocity changes would imply a very low-mass companion, which should be too faint to produce a visible CCF peak. 

Moreover, quadruple systems are intrinsically rare, representing only a small fraction of stellar multiples, particularly in this mass range \citep[less than a few percent; e.g.,][]{2010ApJS..190....1R,2014AJ....147...87T,2023ASPC..534..275O}. Most known quadruples consist of components with comparable masses that likely formed through fragmentation in a common protostellar environment \citep[e.g.,][]{2008MNRAS.389..925T,2025PASP..137i4201S}. A configuration containing a degenerate remnant together with several unevolved low-mass companions would require that the initially most massive star evolved into a WD while the system remained dynamically bound through the giant and mass-loss phases, which makes such a scenario unlikely.

\subsection{Nature of the cool component}

Bringing together the results presented above, the optically dominant component of the central star of K~1-6 is best described as a late-type dwarf exhibiting unusually strong magnetic activity for its inferred stellar parameters. Both the spectroscopic analysis and the position in the \textit{Gaia} colour--magnitude diagram are broadly consistent with a mid-K dwarf, although the star appears somewhat overluminous and likely inflated compared to a single main-sequence object.

The observed inflated radius could be a consequence of past binary interaction, a scenario recently proposed for the class of Abell 35-type objects. These systems, often associated with PNe\footnote{Although the system for which the class is named, Abell 35, has since been shown to be a Str\"omgren sphere rather than a PN \citep{2012A&A...548A.109Z}.}, host late-type companions that show evidence of past mass transfer, including enhanced activity and rapid rotation. Specifically, \citet{2026arXiv260323756B} suggest that these companions may be main-sequence stars currently in a state of thermal inflation due to the energy deposited during a previous accretion phase. {In their models, an accretor contracting post-mass-transfer remains significantly inflated (e.g., $1-3~R_{\odot}$) and rapidly rotating ($P_{\mathrm{rot}} \sim 1-10$ days) for $10^{4}-10^{6}$ years. The inferred radius of $1.5~R_{\odot}$ and the 21.3-day rotational period for the cool component after 1-2 Myr of cooling align reasonably well with these predictions. Furthermore, the intense magnetic activity observed in K 1-6 could potentially boost magnetic braking, thereby increasing the spin-down rate faster than standard models predict. In the context of Abell 35-type objects, the central star of K 1-6 thus likely represents a related evolutionary stage where} interaction with the WD progenitor has altered both the structure and angular momentum of the cool component. 

{The interaction may also have widened the orbit \citep[e.g.,][]{2020ApJ...899..132G,2023A&A...669A..45T,2024ApJ...970L..11H,2025PASP..137j4205Y}, as observed in other post-mass-transfer systems such as barium stars \citep[e.g.,][]{2019A&A...626A.127J,2019A&A...626A.128E}. Furthermore, mass loss during the evolution of the WD progenitor can impart dynamical "kicks" that widen or even unbind wide (tertiary) companions \citep[e.g.,][]{2018MNRAS.480.4884E,2023ApJ...955L..14S,2025ApJ...991..226H}, making this mechanism particularly relevant to the observed wide configuration of the triple system.}

The observational properties of the cool star, including strong chromospheric emission, high X-ray luminosity, large photometric amplitudes, and long-lived spot modulation, also show clear parallels with magnetically active BY~Dra-type systems \citep[e.g.,][]{1973MNRAS.164..343B,1986A&A...165..135R,1997ApJ...478..358D,2022MNRAS.514.4932C}. In such stars, activity is typically driven by rapid rotation and an efficient dynamo. 

There is a remaining tension in this scenario. The projected rotational velocity derived from spectroscopy, $v \sin i \approx 16$~km\,s$^{-1}$, implies a relatively short rotation period. For a canonical K2 dwarf with $R \approx 0.78\,R_\odot$, this would correspond to a rotation period of $\sim$1--3 days (depending on inclination). Even adopting a significantly inflated radius of $\sim$1.5\,R$_\odot$, the expected rotation period would still be only $\sim$2--5 days. This is in clear contrast with the observed photometric period of 21.3 days, which is most naturally interpreted as the stellar rotation period.

Conversely, a rotation period of 21.3 days would imply a much lower equatorial velocity than observed, even when considering a 1.5\,R$_\odot$ inflated star. This discrepancy suggests that either (i) the measured line broadening is not purely rotational and is contaminated by other mechanisms, (ii) the inclination is extremely low (but then the photometric amplitude is too high), or (iii) the star is actually inflated even more. 

A possible explanation related to the last point is that the measured brightness likely represents an average over a highly variable photosphere. In the presented analysis, a simple two-spot model reproduces the observed light curve well, with no spots visible at maximum light. However, this solution is not unique: due to degeneracies in the spot modelling, the stellar surface may host a more complex and extensive spot distribution than captured by the model. As a result, even the observed photometric maximum may not correspond to the true maximum luminosity of the star. In this case, the stellar radius may still be underestimated, and a larger radius would reduce the discrepancy between the observed $v \sin i$ and the rotation period inferred from the photometry. In reality, the observed discrepancy may likely arise from a combination of several of the effects discussed above.

Despite this not being a fully resolved puzzle, the overall phenomenology strongly supports a scenario in which the cool component is a highly magnetically active, spotted star whose properties have likely been shaped by past binary interaction. 

\section{Conclusions}
In this work, we have presented a comprehensive multi-wavelength reanalysis of K 1-6, revealing it to be a far more complex and intriguing object than its original classification as a PN would suggest. Our main conclusions are as follows:
\begin{itemize}
    \item The nebula surrounding K 1-6 is not a surviving PN but rather a photoionised Str\"omgren sphere in the local ISM. 
    \item The central ionising source is a hot WD with a luminosity of only $\approx3.0 L_{\odot}$. This luminosity corresponds to a post-AGB cooling age of approximately 1-2 Myr, a timeframe far beyond the point at which any original PN material would remain structurally or dynamically intact.
    \item We have confirmed that the central star of K 1-6 is a hierarchical triple system. The system consists of an inner binary of a cool component and a WD, likely on an orbit of several hundreds to thousands of days, and a third M2-3V companion in a wide orbit at a projected separation of almost 2000 au, corresponding to an orbital period on the order of tens of thousands of years.
    \item The optically dominant component is a highly active, inflated K-type dwarf exhibiting extreme magnetic activity. This activity is documented through strong H$\alpha$ and Ca II emission, X-ray emission, multiple detected flares, and stable photometric variability (21.3-day period) caused by large, cool spots. Its properties are difficult to reconcile with standard single-star evolution and strongly suggest past binary interaction. Although the system partially still remains a "puzzle" not yet fully solved, it appears closely related to Abell 35-type systems and BY Dra-type binaries.
\end{itemize}

The example of K 1-6, on the one hand, suggests that, had it not been for the high space velocity and photoionisation of the ISM, this system would likely have gone unnoticed, as its original PN has long since dissipated. There might be a population of similar, interesting systems hiding among objects that appear to be normal stars or unremarkable ISM features. On the other hand, the accurate distances provided by \textit{Gaia} are essential for constraining the parameters of central stars and revealing other potentially misclassified PNe. Moreover, in the case of K 1-6, \textit{Gaia} safely resolved the wide third companion specifically because the nebula is faint. In systems with brighter or more complex nebulae, \textit{Gaia} sources often correspond to individual nebular knots or filaments rather than stellar components, making such multiplicity studies significantly more challenging. {Furthermore, tertiary companions are preferentially detectable only in relatively nearby systems with sufficiently wide separations (e.g., $\gtrsim 250$ au for K 1-6, given the $\sim1\arcsec$ angular resolution limit of \textit{Gaia}).}

\begin{acknowledgements}
{We thank the referee for a careful reading of the manuscript and for the constructive comments and suggestions, which helped improve the quality and clarity of this paper.} Based on observations made with the Nordic Optical Telescope, owned in collaboration by the University of Turku and Aarhus University, and operated jointly by Aarhus University, the University of Turku and the University of Oslo, representing Denmark, Finland and Norway, the University of Iceland and Stockholm University at the Observatorio del Roque de los Muchachos, La Palma, Spain, of the Instituto de Astrof\'sica de Canarias. Based on data from the GTC Public Archive at CAB (INTA-CSIC). 

This article includes observations made in the Observatorios de Canarias del IAC with the Telescopio Nazionale Galileo (TNG) operated by the Fundación Galileo Galilei (FGG) of the Istituto Nazionale di Astrofisica (INAF) at the Observatorio del Roque de los Muchachos in the island of La Palma.

This research is based on observations made with the NASA/ESA Hubble Space Telescope obtained from the Space Telescope Science Institute, which is operated by the Association of Universities for Research in Astronomy, Inc., under NASA contract NAS 5–26555. These observations are associated with program 17112.

The research of J.M. was supported by the Czech Science Foundation (GACR) project no. 24-10608O. J.M., T.M., P.G.B., and D.G.R. acknowledge support from the MCIN project PLAtoSOnG (PID2023-146453NB-100).

D.J. and J.G.-R. acknowledge support from the Agencia Estatal de Investigaci\'on del Ministerio de Ciencia, Innovaci\'on y Universidades (MCIU/AEI) under grant ``Nebulosas planetarias como clave para comprender la evoluci\'on de estrellas binarias'' and the European Regional Development Fund (ERDF) with reference PID-2022-136653NA-I00 (DOI:10.13039/501100011033). D.J. and P.S. acknowledge support from the Agencia Estatal de Investigaci\'on del Ministerio de Ciencia, Innovaci\'on y Universidades (MCIU/AEI) under grant ``Revolucionando el conocimiento de la evoluci\'on de estrellas poco masivas'' and the European Union NextGenerationEU/PRTR with reference CNS2023-143910 (DOI:10.13039/501100011033)

A.E. and M.A.M. acknowledge the support of fellowships from ``La Caixa” Foundation (ID 100010434) with fellowship codes LCF/BQ/PI23/11970031 (PI: Ana Escorza) and LCF/BQ/PI23/11970035 (PI: Michael Abdul-Masih).

N.R. is supported by the Deutsche Forschungsgemeinschaft (DFG) through grant RE3915/2-1.

D.G.R. acknowledges support from the Spanish Ministry of Science and Innovation (MICINN) with the Juan de la Cierva fellowship program under contract JDC2022-049054-I.

P.G.B. acknowledges support by the Spanish Ministry of Science, Innovation\&Universities (MCIN) with the Ramón y Cajal fellowship
(RYC-2021-033137-I).

T.M.D. and M.A.P. acknowledge support by the Spanish Ministry of Science via the Plan de Generacion de conocimiento PID2021-124879NB-I00 and PID2024-161863NB-I00. M.A.P. acknowledges support through the Ramón y Cajal grant RYC2022-035388-I, funded by MCIU/AEI/10.13039/501100011033 and FSE+.

S.M. acknowledges support by the Spanish Ministry of Science and Innovation through AEI under the Severo Ochoa Centres of Excellence Programme 2020–2023 (CEX2019-000920-S). S.M. and D.G.R. acknowledge support from the Spanish Ministry of Science and Innovation with the grant no. PID2023-149439NB-C41.

The research of J.M., J.K., M.W., and P.Z. was partially supported by the 
{\sc Cooperatio -- Physics} project of Charles University in Prague.

J.K. acknowledges support from NASA grants NNH22ZDA001N-6152 and 80NSSC24K0638.

The work of H.K. and K.H. was supported by the project RVO:67985815.

\end{acknowledgements}

\bibliographystyle{aa}
\bibliography{lit}

@ARTICLE{Filiz+2024,
       author = {{Filiz}, Semih and {Werner}, Klaus and {Rauch}, Thomas and {Reindl}, Nicole},
        title = "{Spectral evolution of hot hybrid white dwarfs: I. Spectral analysis}",
      journal = {\aap},
         year = 2024,
        month = nov,
       volume = {691},
          eid = {A290},
        pages = {A290},
          doi = {10.1051/0004-6361/202451886},
archivePrefix = {arXiv},
       eprint = {2410.14345},
 primaryClass = {astro-ph.SR},
       adsurl = {https://ui.adsabs.harvard.edu/abs/2024A&A...691A.290F}
}

@ARTICLE{Heber+2018,
       author = {{Heber}, Ulrich and {Irrgang}, Andreas and {Schaffenroth}, Johannes},
        title = "{Spectral energy distributions and colours of hot subluminous stars}",
      journal = {Open Astronomy},
         year = 2018,
        month = feb,
       volume = {27},
       number = {1},
        pages = {35-43},
          doi = {10.1515/astro-2018-0008},
archivePrefix = {arXiv},
       eprint = {1712.06546},
 primaryClass = {astro-ph.SR},
       adsurl = {https://ui.adsabs.harvard.edu/abs/2018OAst...27...35H}
}

@ARTICLE{Irrgang+2021,
       author = {{Irrgang}, A. and {Geier}, S. and {Heber}, U. and {Kupfer}, T. and {El-Badry}, K. and {Bloemen}, S.},
        title = "{A proto-helium white dwarf stripped by a substellar companion via common-envelope ejection. Uncovering the true nature of a candidate hypervelocity B-type star}",
      journal = {\aap},
         year = 2021,
        month = jun,
       volume = {650},
          eid = {A102},
        pages = {A102},
          doi = {10.1051/0004-6361/202038757},
archivePrefix = {arXiv},
       eprint = {2007.03350},
 primaryClass = {astro-ph.SR},
       adsurl = {https://ui.adsabs.harvard.edu/abs/2021A&A...650A.102I}
}

@ARTICLE{Fitzpatrick+2019,
       author = {{Fitzpatrick}, E.~L. and {Massa}, Derck and {Gordon}, Karl D. and {Bohlin}, Ralph and {Clayton}, Geoffrey C.},
        title = "{An Analysis of the Shapes of Interstellar Extinction Curves. VII. Milky Way Spectrophotometric Optical-through-ultraviolet Extinction and Its R-dependence}",
      journal = {\apj},
         year = 2019,
        month = dec,
       volume = {886},
       number = {2},
          eid = {108},
        pages = {108},
          doi = {10.3847/1538-4357/ab4c3a},
archivePrefix = {arXiv},
       eprint = {1910.08852},
 primaryClass = {astro-ph.GA},
       adsurl = {https://ui.adsabs.harvard.edu/abs/2019ApJ...886..108F}
}

@ARTICLE{Husser+2013,
       author = {{Husser}, T. -O. and {Wende-von Berg}, S. and {Dreizler}, S. and {Homeier}, D. and {Reiners}, A. and {Barman}, T. and {Hauschildt}, P.~H.},
        title = "{A new extensive library of PHOENIX stellar atmospheres and synthetic spectra}",
      journal = {\aap},
         year = 2013,
        month = may,
       volume = {553},
          eid = {A6},
        pages = {A6},
          doi = {10.1051/0004-6361/201219058},
archivePrefix = {arXiv},
       eprint = {1303.5632},
 primaryClass = {astro-ph.SR},
       adsurl = {https://ui.adsabs.harvard.edu/abs/2013A&A...553A...6H}
}

@ARTICLE{Reindl+2023,
       author = {{Reindl}, Nicole and {Islami}, Ramazan and {Werner}, Klaus and {Kepler}, S.~O. and {Pritzkuleit}, Max and {Dawson}, Harry and {Dorsch}, Matti and {Istrate}, Alina and {Pelisoli}, Ingrid and {Geier}, Stephan and {Uzundag}, Murat and {Provencal}, Judith and {Justham}, Stephen},
        title = "{The bright blue side of the night sky: Spectroscopic survey of bright and hot (pre-) white dwarfs}",
      journal = {\aap},
         year = 2023,
        month = sep,
       volume = {677},
          eid = {A29},
        pages = {A29},
          doi = {10.1051/0004-6361/202346865},
archivePrefix = {arXiv},
       eprint = {2307.03721},
 primaryClass = {astro-ph.SR},
       adsurl = {https://urldefense.com/v3/__https://ui.adsabs.harvard.edu/abs/2023A&A...677A..29R__;!!CrWY41Z8OgsX0i-WU-0LuAcUu2o!2c0Fajx8aucDgVoDIZhbyXvftoPOkcNkCoBuZ_1a4-DkqkTAm_rIzczATO1FUuEbN9iEQzZmt5R-dvpP$ }
}

@ARTICLE{Renedo+2010,
       author = {{Renedo}, I. and {Althaus}, L.~G. and {Miller Bertolami}, M.~M. and
         {Romero}, A.~D. and {C{\'o}rsico}, A.~H. and {Rohrmann}, R.~D. and
         {Garc{\'\i}a-Berro}, E.},
        title = "{New Cooling Sequences for Old White Dwarfs}",
      journal = {\apj},
         year = 2010,
        month = jul,
       volume = {717},
       number = {1},
        pages = {183-195},
          doi = {10.1088/0004-637X/717/1/183},
archivePrefix = {arXiv},
       eprint = {1005.2170},
 primaryClass = {astro-ph.SR},
       adsurl = {https://ui.adsabs.harvard.edu/abs/2010ApJ...717..183R}
}

@ARTICLE{1962BAICz..13..120K,
       author = {{Kohoutek}, L.},
        title = "{New planetary nebulae}",
      journal = {Bulletin of the Astronomical Institutes of Czechoslovakia},
         year = 1962,
        month = jan,
       volume = {13},
        pages = {120},
       adsurl = {https://ui.adsabs.harvard.edu/abs/1962BAICz..13..120K}
}

@ARTICLE{2017ApJS..230...24B,
       author = {{Bianchi}, Luciana and {Shiao}, Bernie and {Thilker}, David},
        title = "{Revised Catalog of GALEX Ultraviolet Sources. I. The All-Sky Survey: GUVcat\_AIS}",
      journal = {\apjs},
         year = 2017,
        month = jun,
       volume = {230},
       number = {2},
          eid = {24},
        pages = {24},
          doi = {10.3847/1538-4365/aa7053},
archivePrefix = {arXiv},
       eprint = {1704.05903},
 primaryClass = {astro-ph.GA},
       adsurl = {https://ui.adsabs.harvard.edu/abs/2017ApJS..230...24B}
}

@ARTICLE{2005LRSP....2....8B,
       author = {{Berdyugina}, Svetlana V.},
        title = "{Starspots: A Key to the Stellar Dynamo}",
      journal = {Living Reviews in Solar Physics},
         year = 2005,
        month = dec,
       volume = {2},
       number = {1},
          eid = {8},
        pages = {8},
          doi = {10.12942/lrsp-2005-8},
       adsurl = {https://ui.adsabs.harvard.edu/abs/2005LRSP....2....8B}
}

@ARTICLE{2017MNRAS.472.1618G,
       author = {{Giles}, Helen A.~C. and {Collier Cameron}, Andrew and {Haywood}, Rapha{\"e}lle D.},
        title = "{A Kepler study of starspot lifetimes with respect to light-curve amplitude and spectral type}",
      journal = {\mnras},
         year = 2017,
        month = dec,
       volume = {472},
       number = {2},
        pages = {1618-1627},
          doi = {10.1093/mnras/stx1931},
archivePrefix = {arXiv},
       eprint = {1707.08583},
 primaryClass = {astro-ph.SR},
       adsurl = {https://ui.adsabs.harvard.edu/abs/2017MNRAS.472.1618G}
}

@ARTICLE{2021A&A...656A.110C,
       author = {{Chornay}, N. and {Walton}, N.~A.},
        title = "{One star, two star, red star, blue star: an updated planetary nebula central star distance catalogue from Gaia EDR3}",
      journal = {\aap},
         year = 2021,
        month = dec,
       volume = {656},
          eid = {A110},
        pages = {A110},
          doi = {10.1051/0004-6361/202142008},
archivePrefix = {arXiv},
       eprint = {2102.13654},
 primaryClass = {astro-ph.SR},
       adsurl = {https://ui.adsabs.harvard.edu/abs/2021A&A...656A.110C}
}

@BOOK{2019ibfe.book.....B,
       author = {{Boffin}, Henri M.~J. and {Jones}, David},
        title = "{The Importance of Binaries in the Formation and Evolution of Planetary Nebulae}",
         year = 2019,
          doi = {10.1007/978-3-030-25059-1},
       adsurl = {https://ui.adsabs.harvard.edu/abs/2019ibfe.book.....B}
}

@ARTICLE{2017NatAs...1E.117J,
       author = {{Jones}, David and {Boffin}, Henri M.~J.},
        title = "{Binary stars as the key to understanding planetary nebulae}",
      journal = {Nature Astronomy},
         year = 2017,
        month = may,
       volume = {1},
          eid = {0117},
        pages = {0117},
          doi = {10.1038/s41550-017-0117},
archivePrefix = {arXiv},
       eprint = {1705.00283},
 primaryClass = {astro-ph.SR},
       adsurl = {https://ui.adsabs.harvard.edu/abs/2017NatAs...1E.117J}
}

@ARTICLE{2021A&A...656A..51G,
       author = {{Gonz{\'a}lez-Santamar{\'\i}a}, I. and {Manteiga}, M. and {Manchado}, A. and {Ulla}, A. and {Dafonte}, C. and {L{\'o}pez Varela}, P.},
        title = "{Planetary nebulae in Gaia EDR3: Central star identification, properties, and binarity}",
      journal = {\aap},
         year = 2021,
        month = dec,
       volume = {656},
          eid = {A51},
        pages = {A51},
          doi = {10.1051/0004-6361/202141916},
archivePrefix = {arXiv},
       eprint = {2109.12114},
 primaryClass = {astro-ph.GA},
       adsurl = {https://ui.adsabs.harvard.edu/abs/2021A&A...656A..51G}
}

@ARTICLE{2025MNRAS.543.3035C,
       author = {{Csukai}, Alexander and {Zijlstra}, Albert A. and {McDonald}, Iain and {De Marco}, Orsola},
        title = "{Central-star extinctions towards planetary nebulae}",
      journal = {\mnras},
         year = 2025,
        month = nov,
       volume = {543},
       number = {3},
        pages = {3035-3054},
          doi = {10.1093/mnras/staf1552},
archivePrefix = {arXiv},
       eprint = {2509.10621},
 primaryClass = {astro-ph.SR},
       adsurl = {https://ui.adsabs.harvard.edu/abs/2025MNRAS.543.3035C}
}

@ARTICLE{1988AJ.....96..997K,
       author = {{Kwitter}, Karen B. and {Jacoby}, George H. and {Lydon}, Thomas J.},
        title = "{Identifications of Faint Central Stars in Extended, Low-Surface-Brightness Planetary Nebulae}",
      journal = {\aj},
         year = 1988,
        month = sep,
       volume = {96},
        pages = {997},
          doi = {10.1086/114859},
       adsurl = {https://ui.adsabs.harvard.edu/abs/1988AJ.....96..997K}
}

@ARTICLE{2008OEJV...88....1U,
       author = {{Usatov}, Maxim and {Nosulchik}, Artem},
        title = "{108 New Variable Stars in the NSVS Database}",
      journal = {Open European Journal on Variable Stars},
         year = 2008,
        month = may,
       volume = {0088},
       number = {1},
        pages = {1},
       adsurl = {https://ui.adsabs.harvard.edu/abs/2008OEJV...88....1U}
}

@ARTICLE{2023MNRAS.520..773R,
       author = {{Ritter}, Andreas and {Parker}, Q.~A. and {Sabin}, L. and {Le D{\^u}}, P. and {Mulato}, L. and {Patchick}, D.},
        title = "{Grantecan spectroscopic observations and confirmations of planetary nebulae candidates in the Northern Galactic Plane}",
      journal = {\mnras},
         year = 2023,
        month = mar,
       volume = {520},
       number = {1},
        pages = {773-781},
          doi = {10.1093/mnras/stac2896},
archivePrefix = {arXiv},
       eprint = {2210.07581},
 primaryClass = {astro-ph.SR},
       adsurl = {https://ui.adsabs.harvard.edu/abs/2023MNRAS.520..773R}
}

@ARTICLE{2017PASP..129j4502K,
       author = {{Kochanek}, C.~S. and {Shappee}, B.~J. and {Stanek}, K.~Z. and {Holoien}, T.~W. -S. and {Thompson}, Todd A. and {Prieto}, J.~L. and {Dong}, Subo and {Shields}, J.~V. and {Will}, D. and {Britt}, C. and {Perzanowski}, D. and {Pojma{\'n}ski}, G.},
        title = "{The All-Sky Automated Survey for Supernovae (ASAS-SN) Light Curve Server v1.0}",
      journal = {\pasp},
         year = 2017,
        month = oct,
       volume = {129},
       number = {980},
        pages = {104502},
          doi = {10.1088/1538-3873/aa80d9},
archivePrefix = {arXiv},
       eprint = {1706.07060},
 primaryClass = {astro-ph.SR},
       adsurl = {https://ui.adsabs.harvard.edu/abs/2017PASP..129j4502K}
}

@ARTICLE{2014ApJ...788...48S,
       author = {{Shappee}, B.~J. and {Prieto}, J.~L. and {Grupe}, D. and {Kochanek}, C.~S. and {Stanek}, K.~Z. and {De Rosa}, G. and {Mathur}, S. and {Zu}, Y. and {Peterson}, B.~M. and {Pogge}, R.~W. and {Komossa}, S. and {Im}, M. and {Jencson}, J. and {Holoien}, T.~W. -S. and {Basu}, U. and {Beacom}, J.~F. and {Szczygie{\l}}, D.~M. and {Brimacombe}, J. and {Adams}, S. and {Campillay}, A. and {Choi}, C. and {Contreras}, C. and {Dietrich}, M. and {Dubberley}, M. and {Elphick}, M. and {Foale}, S. and {Giustini}, M. and {Gonzalez}, C. and {Hawkins}, E. and {Howell}, D.~A. and {Hsiao}, E.~Y. and {Koss}, M. and {Leighly}, K.~M. and {Morrell}, N. and {Mudd}, D. and {Mullins}, D. and {Nugent}, J.~M. and {Parrent}, J. and {Phillips}, M.~M. and {Pojmanski}, G. and {Rosing}, W. and {Ross}, R. and {Sand}, D. and {Terndrup}, D.~M. and {Valenti}, S. and {Walker}, Z. and {Yoon}, Y.},
        title = "{The Man behind the Curtain: X-Rays Drive the UV through NIR Variability in the 2013 Active Galactic Nucleus Outburst in NGC 2617}",
      journal = {\apj},
         year = 2014,
        month = jun,
       volume = {788},
       number = {1},
          eid = {48},
        pages = {48},
          doi = {10.1088/0004-637X/788/1/48},
archivePrefix = {arXiv},
       eprint = {1310.2241},
 primaryClass = {astro-ph.HE},
       adsurl = {https://ui.adsabs.harvard.edu/abs/2014ApJ...788...48S}
}

@ARTICLE{2019PASP..131a8003M,
       author = {{Masci}, Frank J. and {Laher}, Russ R. and {Rusholme}, Ben and {Shupe}, David L. and {Groom}, Steven and {Surace}, Jason and {Jackson}, Edward and {Monkewitz}, Serge and {Beck}, Ron and {Flynn}, David and {Terek}, Scott and {Landry}, Walter and {Hacopians}, Eugean and {Desai}, Vandana and {Howell}, Justin and {Brooke}, Tim and {Imel}, David and {Wachter}, Stefanie and {Ye}, Quan-Zhi and {Lin}, Hsing-Wen and {Cenko}, S. Bradley and {Cunningham}, Virginia and {Rebbapragada}, Umaa and {Bue}, Brian and {Miller}, Adam A. and {Mahabal}, Ashish and {Bellm}, Eric C. and {Patterson}, Maria T. and {Juri{\'c}}, Mario and {Golkhou}, V. Zach and {Ofek}, Eran O. and {Walters}, Richard and {Graham}, Matthew and {Kasliwal}, Mansi M. and {Dekany}, Richard G. and {Kupfer}, Thomas and {Burdge}, Kevin and {Cannella}, Christopher B. and {Barlow}, Tom and {Van Sistine}, Angela and {Giomi}, Matteo and {Fremling}, Christoffer and {Blagorodnova}, Nadejda and {Levitan}, David and {Riddle}, Reed and {Smith}, Roger M. and {Helou}, George and {Prince}, Thomas A. and {Kulkarni}, Shrinivas R.},
        title = "{The Zwicky Transient Facility: Data Processing, Products, and Archive}",
      journal = {\pasp},
         year = 2019,
        month = jan,
       volume = {131},
       number = {995},
        pages = {018003},
          doi = {10.1088/1538-3873/aae8ac},
archivePrefix = {arXiv},
       eprint = {1902.01872},
 primaryClass = {astro-ph.IM},
       adsurl = {https://ui.adsabs.harvard.edu/abs/2019PASP..131a8003M}
}

@ARTICLE{2015JATIS...1a4003R,
       author = {{Ricker}, George R. and {Winn}, Joshua N. and {Vanderspek}, Roland and {Latham}, David W. and {Bakos}, G{\'a}sp{\'a}r {\'A}. and {Bean}, Jacob L. and {Berta-Thompson}, Zachory K. and {Brown}, Timothy M. and {Buchhave}, Lars and {Butler}, Nathaniel R. and {Butler}, R. Paul and {Chaplin}, William J. and {Charbonneau}, David and {Christensen-Dalsgaard}, J{\o}rgen and {Clampin}, Mark and {Deming}, Drake and {Doty}, John and {De Lee}, Nathan and {Dressing}, Courtney and {Dunham}, Edward W. and {Endl}, Michael and {Fressin}, Francois and {Ge}, Jian and {Henning}, Thomas and {Holman}, Matthew J. and {Howard}, Andrew W. and {Ida}, Shigeru and {Jenkins}, Jon M. and {Jernigan}, Garrett and {Johnson}, John Asher and {Kaltenegger}, Lisa and {Kawai}, Nobuyuki and {Kjeldsen}, Hans and {Laughlin}, Gregory and {Levine}, Alan M. and {Lin}, Douglas and {Lissauer}, Jack J. and {MacQueen}, Phillip and {Marcy}, Geoffrey and {McCullough}, Peter R. and {Morton}, Timothy D. and {Narita}, Norio and {Paegert}, Martin and {Palle}, Enric and {Pepe}, Francesco and {Pepper}, Joshua and {Quirrenbach}, Andreas and {Rinehart}, Stephen A. and {Sasselov}, Dimitar and {Sato}, Bun'ei and {Seager}, Sara and {Sozzetti}, Alessandro and {Stassun}, Keivan G. and {Sullivan}, Peter and {Szentgyorgyi}, Andrew and {Torres}, Guillermo and {Udry}, Stephane and {Villasenor}, Joel},
        title = "{Transiting Exoplanet Survey Satellite (TESS)}",
      journal = {Journal of Astronomical Telescopes, Instruments, and Systems},
         year = 2015,
        month = jan,
       volume = {1},
          eid = {014003},
        pages = {014003},
          doi = {10.1117/1.JATIS.1.1.014003},
       adsurl = {https://ui.adsabs.harvard.edu/abs/2015JATIS...1a4003R}
}

@ARTICLE{2023Ap&SS.368...34A,
       author = {{Ali}, A. and {Khalil}, J.~M. and {Mindil}, A.},
        title = "{Detection of wide binary and multiple nuclei of planetary nebulae using the Gaia DR3}",
      journal = {\apss},
         year = 2023,
        month = apr,
       volume = {368},
       number = {4},
          eid = {34},
        pages = {34},
          doi = {10.1007/s10509-023-04180-8},
archivePrefix = {arXiv},
       eprint = {2303.06191},
 primaryClass = {astro-ph.SR},
       adsurl = {https://ui.adsabs.harvard.edu/abs/2023Ap&SS.368...34A}
}

@ARTICLE{2023A&A...674A...1G,
       author = {{Gaia Collaboration} and {Vallenari}, A. and {Brown}, A.~G.~A. and {Prusti}, T. and {de Bruijne}, J.~H.~J. and {Arenou}, F. and {Babusiaux}, C. and {Biermann}, M. and {Creevey}, O.~L. and {Ducourant}, C. and {Evans}, D.~W. and {Eyer}, L. and {Guerra}, R. and {Hutton}, A. and {Jordi}, C. and {Klioner}, S.~A. and {Lammers}, U.~L. and {Lindegren}, L. and {Luri}, X. and {Mignard}, F. and {Panem}, C. and {Pourbaix}, D. and {Randich}, S. and {Sartoretti}, P. and {Soubiran}, C. and {Tanga}, P. and {Walton}, N.~A. and {Bailer-Jones}, C.~A.~L. and {Bastian}, U. and {Drimmel}, R. and {Jansen}, F. and {Katz}, D. and {Lattanzi}, M.~G. and {van Leeuwen}, F. and {Bakker}, J. and {Cacciari}, C. and {Casta{\~n}eda}, J. and {De Angeli}, F. and {Fabricius}, C. and {Fouesneau}, M. and {Fr{\'e}mat}, Y. and {Galluccio}, L. and {Guerrier}, A. and {Heiter}, U. and {Masana}, E. and {Messineo}, R. and {Mowlavi}, N. and {Nicolas}, C. and {Nienartowicz}, K. and {Pailler}, F. and {Panuzzo}, P. and {Riclet}, F. and {Roux}, W. and {Seabroke}, G.~M. and {Sordo}, R. and {Th{\'e}venin}, F. and {Gracia-Abril}, G. and {Portell}, J. and {Teyssier}, D. and {Altmann}, M. and {Andrae}, R. and {Audard}, M. and {Bellas-Velidis}, I. and {Benson}, K. and {Berthier}, J. and {Blomme}, R. and {Burgess}, P.~W. and {Busonero}, D. and {Busso}, G. and {C{\'a}novas}, H. and {Carry}, B. and {Cellino}, A. and {Cheek}, N. and {Clementini}, G. and {Damerdji}, Y. and {Davidson}, M. and {de Teodoro}, P. and {Nu{\~n}ez Campos}, M. and {Delchambre}, L. and {Dell'Oro}, A. and {Esquej}, P. and {Fern{\'a}ndez-Hern{\'a}ndez}, J. and {Fraile}, E. and {Garabato}, D. and {Garc{\'\i}a-Lario}, P. and {Gosset}, E. and {Haigron}, R. and {Halbwachs}, J. -L. and {Hambly}, N.~C. and {Harrison}, D.~L. and {Hern{\'a}ndez}, J. and {Hestroffer}, D. and {Hodgkin}, S.~T. and {Holl}, B. and {Jan{\ss}en}, K. and {Jevardat de Fombelle}, G. and {Jordan}, S. and {Krone-Martins}, A. and {Lanzafame}, A.~C. and {L{\"o}ffler}, W. and {Marchal}, O. and {Marrese}, P.~M. and {Moitinho}, A. and {Muinonen}, K. and {Osborne}, P. and {Pancino}, E. and {Pauwels}, T. and {Recio-Blanco}, A. and {Reyl{\'e}}, C. and {Riello}, M. and {Rimoldini}, L. and {Roegiers}, T. and {Rybizki}, J. and {Sarro}, L.~M. and {Siopis}, C. and {Smith}, M. and {Sozzetti}, A. and {Utrilla}, E. and {van Leeuwen}, M. and {Abbas}, U. and {{\'A}brah{\'a}m}, P. and {Abreu Aramburu}, A. and {Aerts}, C. and {Aguado}, J.~J. and {Ajaj}, M. and {Aldea-Montero}, F. and {Altavilla}, G. and {{\'A}lvarez}, M.~A. and {Alves}, J. and {Anders}, F. and {Anderson}, R.~I. and {Anglada Varela}, E. and {Antoja}, T. and {Baines}, D. and {Baker}, S.~G. and {Balaguer-N{\'u}{\~n}ez}, L. and {Balbinot}, E. and {Balog}, Z. and {Barache}, C. and {Barbato}, D. and {Barros}, M. and {Barstow}, M.~A. and {Bartolom{\'e}}, S. and {Bassilana}, J. -L. and {Bauchet}, N. and {Becciani}, U. and {Bellazzini}, M. and {Berihuete}, A. and {Bernet}, M. and {Bertone}, S. and {Bianchi}, L. and {Binnenfeld}, A. and {Blanco-Cuaresma}, S. and {Blazere}, A. and {Boch}, T. and {Bombrun}, A. and {Bossini}, D. and {Bouquillon}, S. and {Bragaglia}, A. and {Bramante}, L. and {Breedt}, E. and {Bressan}, A. and {Brouillet}, N. and {Brugaletta}, E. and {Bucciarelli}, B. and {Burlacu}, A. and {Butkevich}, A.~G. and {Buzzi}, R. and {Caffau}, E. and {Cancelliere}, R. and {Cantat-Gaudin}, T. and {Carballo}, R. and {Carlucci}, T. and {Carnerero}, M.~I. and {Carrasco}, J.~M. and {Casamiquela}, L. and {Castellani}, M. and {Castro-Ginard}, A. and {Chaoul}, L. and {Charlot}, P. and {Chemin}, L. and {Chiaramida}, V. and {Chiavassa}, A. and {Chornay}, N. and {Comoretto}, G. and {Contursi}, G. and {Cooper}, W.~J. and {Cornez}, T. and {Cowell}, S. and {Crifo}, F. and {Cropper}, M. and {Crosta}, M. and {Crowley}, C. and {Dafonte}, C. and {Dapergolas}, A. and {David}, M. and {David}, P. and {de Laverny}, P. and {De Luise}, F. and {De March}, R.},
        title = "{Gaia Data Release 3. Summary of the content and survey properties}",
      journal = {\aap},
         year = 2023,
        month = jun,
       volume = {674},
          eid = {A1},
        pages = {A1},
          doi = {10.1051/0004-6361/202243940},
archivePrefix = {arXiv},
       eprint = {2208.00211},
 primaryClass = {astro-ph.GA},
       adsurl = {https://ui.adsabs.harvard.edu/abs/2023A&A...674A...1G}
}

@ARTICLE{2023A&A...674A...9H,
       author = {{Halbwachs}, Jean-Louis and {Pourbaix}, Dimitri and {Arenou}, Fr{\'e}d{\'e}ric and {Galluccio}, Laurent and {Guillout}, Patrick and {Bauchet}, Nathalie and {Marchal}, Olivier and {Sadowski}, Gilles and {Teyssier}, David},
        title = "{Gaia Data Release 3. Astrometric binary star processing}",
      journal = {\aap},
         year = 2023,
        month = jun,
       volume = {674},
          eid = {A9},
        pages = {A9},
          doi = {10.1051/0004-6361/202243969},
archivePrefix = {arXiv},
       eprint = {2206.05726},
 primaryClass = {astro-ph.SR},
       adsurl = {https://ui.adsabs.harvard.edu/abs/2023A&A...674A...9H}
}

@ARTICLE{2020PASP..132h5002S,
       author = {{Smith}, K.~W. and {Smartt}, S.~J. and {Young}, D.~R. and {Tonry}, J.~L. and {Denneau}, L. and {Flewelling}, H. and {Heinze}, A.~N. and {Weiland}, H.~J. and {Stalder}, B. and {Rest}, A. and {Stubbs}, C.~W. and {Anderson}, J.~P. and {Chen}, T. -W. and {Clark}, P. and {Do}, A. and {F{\"o}rster}, F. and {Fulton}, M. and {Gillanders}, J. and {McBrien}, O.~R. and {O'Neill}, D. and {Srivastav}, S. and {Wright}, D.~E.},
        title = "{Design and Operation of the ATLAS Transient Science Server}",
      journal = {\pasp},
         year = 2020,
        month = aug,
       volume = {132},
       number = {1014},
          eid = {085002},
        pages = {085002},
          doi = {10.1088/1538-3873/ab936e},
archivePrefix = {arXiv},
       eprint = {2003.09052},
 primaryClass = {astro-ph.IM},
       adsurl = {https://ui.adsabs.harvard.edu/abs/2020PASP..132h5002S}
}

@ARTICLE{2018PASP..130f4505T,
       author = {{Tonry}, J.~L. and {Denneau}, L. and {Heinze}, A.~N. and {Stalder}, B. and {Smith}, K.~W. and {Smartt}, S.~J. and {Stubbs}, C.~W. and {Weiland}, H.~J. and {Rest}, A.},
        title = "{ATLAS: A High-cadence All-sky Survey System}",
      journal = {\pasp},
         year = 2018,
        month = jun,
       volume = {130},
       number = {988},
        pages = {064505},
          doi = {10.1088/1538-3873/aabadf},
archivePrefix = {arXiv},
       eprint = {1802.00879},
 primaryClass = {astro-ph.IM},
       adsurl = {https://ui.adsabs.harvard.edu/abs/2018PASP..130f4505T}
}

@ARTICLE{2021TNSAN...7....1S,
       author = {{Shingles}, L. and {Smith}, K.~W. and {Young}, D.~R. and {Smartt}, S.~J. and {Tonry}, J. and {Denneau}, L. and {Heinze}, A. and {Weiland}, H. and {Flewelling}, H. and {Stalder}, B. and {Clocchiatti}, A. and {F{\"o}rster}, F. and {Pignata}, G. and {Rest}, A. and {Anderson}, J. and {Stubbs}, C. and {Erasmus}, N.},
        title = "{Release of the ATLAS Forced Photometry server for public use}",
      journal = {Transient Name Server AstroNote},
         year = 2021,
        month = jan,
       volume = {7},
        pages = {1-7},
       adsurl = {https://ui.adsabs.harvard.edu/abs/2021TNSAN...7....1S}
}

@ARTICLE{2013ApJS..208....9P,
       author = {{Pecaut}, Mark J. and {Mamajek}, Eric E.},
        title = "{Intrinsic Colors, Temperatures, and Bolometric Corrections of Pre-main-sequence Stars}",
      journal = {\apjs},
         year = 2013,
        month = sep,
       volume = {208},
       number = {1},
          eid = {9},
        pages = {9},
          doi = {10.1088/0067-0049/208/1/9},
archivePrefix = {arXiv},
       eprint = {1307.2657},
 primaryClass = {astro-ph.SR},
       adsurl = {https://ui.adsabs.harvard.edu/abs/2013ApJS..208....9P}
}

@ARTICLE{2010ApJS..190....1R,
       author = {{Raghavan}, Deepak and {McAlister}, Harold A. and {Henry}, Todd J. and {Latham}, David W. and {Marcy}, Geoffrey W. and {Mason}, Brian D. and {Gies}, Douglas R. and {White}, Russel J. and {ten Brummelaar}, Theo A.},
        title = "{A Survey of Stellar Families: Multiplicity of Solar-type Stars}",
      journal = {\apjs},
         year = 2010,
        month = sep,
       volume = {190},
       number = {1},
        pages = {1-42},
          doi = {10.1088/0067-0049/190/1/1},
archivePrefix = {arXiv},
       eprint = {1007.0414},
 primaryClass = {astro-ph.SR},
       adsurl = {https://ui.adsabs.harvard.edu/abs/2010ApJS..190....1R}
}

@ARTICLE{2014AJ....147...87T,
       author = {{Tokovinin}, Andrei},
        title = "{From Binaries to Multiples. II. Hierarchical Multiplicity of F and G Dwarfs}",
      journal = {\aj},
         year = 2014,
        month = apr,
       volume = {147},
       number = {4},
          eid = {87},
        pages = {87},
          doi = {10.1088/0004-6256/147/4/87},
archivePrefix = {arXiv},
       eprint = {1401.6827},
 primaryClass = {astro-ph.SR},
       adsurl = {https://ui.adsabs.harvard.edu/abs/2014AJ....147...87T}
}

@INPROCEEDINGS{2023ASPC..534..275O,
       author = {{Offner}, S.~S.~R. and {Moe}, M. and {Kratter}, K.~M. and {Sadavoy}, S.~I. and {Jensen}, E.~L.~N. and {Tobin}, J.~J.},
        title = "{The Origin and Evolution of Multiple Star Systems}",
    booktitle = {Protostars and Planets VII},
         year = 2023,
       editor = {{Inutsuka}, S. and {Aikawa}, Y. and {Muto}, T. and {Tomida}, K. and {Tamura}, M.},
       series = {Astronomical Society of the Pacific Conference Series},
       volume = {534},
        month = jul,
        pages = {275},
          doi = {10.48550/arXiv.2203.10066},
archivePrefix = {arXiv},
       eprint = {2203.10066},
 primaryClass = {astro-ph.SR},
       adsurl = {https://ui.adsabs.harvard.edu/abs/2023ASPC..534..275O}
}

@ARTICLE{2008MNRAS.389..925T,
       author = {{Tokovinin}, A.},
        title = "{Comparative statistics and origin of triple and quadruple stars}",
      journal = {\mnras},
         year = 2008,
        month = sep,
       volume = {389},
       number = {2},
        pages = {925-938},
          doi = {10.1111/j.1365-2966.2008.13613.x},
archivePrefix = {arXiv},
       eprint = {0806.3263},
 primaryClass = {astro-ph},
       adsurl = {https://ui.adsabs.harvard.edu/abs/2008MNRAS.389..925T}
}

@ARTICLE{2025PASP..137i4201S,
       author = {{Shariat}, Cheyanne and {El-Badry}, Kareem and {Naoz}, Smadar},
        title = "{10,000 Resolved Triples from Gaia: Empirical Constraints on Triple Star Populations}",
      journal = {\pasp},
         year = 2025,
        month = sep,
       volume = {137},
       number = {9},
          eid = {094201},
        pages = {094201},
          doi = {10.1088/1538-3873/adfb30},
archivePrefix = {arXiv},
       eprint = {2506.16513},
 primaryClass = {astro-ph.SR},
       adsurl = {https://ui.adsabs.harvard.edu/abs/2025PASP..137i4201S}
}

@ARTICLE{2026arXiv260323756B,
       author = {{Bhattacharjee}, Soumyadeep and {El-Badry}, Kareem and {Fuller}, Jim and {Shariat}, Cheyanne and {Yamaguchi}, Natsuko},
        title = "{Thermally inflated accretors in post-mass transfer binaries: Abell 35 and its class revisited}",
      journal = {arXiv e-prints},
         year = 2026,
        month = mar,
          eid = {arXiv:2603.23756},
        pages = {arXiv:2603.23756},
          doi = {10.48550/arXiv.2603.23756},
archivePrefix = {arXiv},
       eprint = {2603.23756},
 primaryClass = {astro-ph.SR},
       adsurl = {https://ui.adsabs.harvard.edu/abs/2026arXiv260323756B}
}

@ARTICLE{2019A&A...626A.128E,
       author = {{Escorza}, A. and {Karinkuzhi}, D. and {Jorissen}, A. and {Siess}, L. and {Van Winckel}, H. and {Pourbaix}, D. and {Johnston}, C. and {Miszalski}, B. and {Oomen}, G.-M. and {Abdul-Masih}, M. and {Boffin}, H.~M.~J. and {North}, P. and {Manick}, R. and {Shetye}, S. and {Miko{\l}ajewska}, J.},
        title = "{Barium and related stars, and their white-dwarf companions. II. Main-sequence and subgiant starss}",
      journal = {\aap},
         year = 2019,
        month = jun,
       volume = {626},
          eid = {A128},
        pages = {A128},
          doi = {10.1051/0004-6361/201935390},
archivePrefix = {arXiv},
       eprint = {1904.04095},
 primaryClass = {astro-ph.SR},
       adsurl = {https://ui.adsabs.harvard.edu/abs/2019A&A...626A.128E}
}

@ARTICLE{2019A&A...626A.127J,
       author = {{Jorissen}, A. and {Boffin}, H.~M.~J. and {Karinkuzhi}, D. and {Van Eck}, S. and {Escorza}, A. and {Shetye}, S. and {Van Winckel}, H.},
        title = "{Barium and related stars, and their white-dwarf companions. I. Giant stars}",
      journal = {\aap},
         year = 2019,
        month = jun,
       volume = {626},
          eid = {A127},
        pages = {A127},
          doi = {10.1051/0004-6361/201834630},
archivePrefix = {arXiv},
       eprint = {1904.03975},
 primaryClass = {astro-ph.SR},
       adsurl = {https://ui.adsabs.harvard.edu/abs/2019A&A...626A.127J}
}

@ARTICLE{1986A&A...165..135R,
       author = {{Rodono}, M. and {Cutispoto}, G. and {Pazzani}, V. and {Catalano}, S. and {Byrne}, P.~B. and {Doyle}, J.~G. and {Butler}, C.~J. and {Andrews}, A.~D. and {Blanco}, C. and {Marilli}, E. and {Linsky}, J.~L. and {Scaltriti}, F. and {Busso}, M. and {Cellino}, A. and {Hopkins}, J.~L. and {Okazaki}, A. and {Hayashi}, S.~S. and {Zeilik}, M. and {Helston}, R. and {Henson}, G. and {Smith}, P. and {Simon}, T.},
        title = "{Rotational modulation and flares on RS CVn and BY Dra-type stars. I. Photometry and SPOT models for BY Dra, AU Mic, AR Lac, II Peg and V711 Tau (=HR 1099).}",
      journal = {\aap},
         year = 1986,
        month = sep,
       volume = {165},
        pages = {135-156},
       adsurl = {https://ui.adsabs.harvard.edu/abs/1986A&A...165..135R}
}

@ARTICLE{1997ApJ...478..358D,
       author = {{Dempsey}, Robert C. and {Linsky}, Jeffrey L. and {Fleming}, Thomas A. and {Schmitt}, J.~H.~M.~M.},
        title = "{The ROSAT All-Sky Survey of Active Binary Coronae. III. Quiescent Coronal Properties for the BY Draconis-Type Binaries}",
      journal = {\apj},
         year = 1997,
        month = mar,
       volume = {478},
       number = {1},
        pages = {358-366},
          doi = {10.1086/303786},
       adsurl = {https://ui.adsabs.harvard.edu/abs/1997ApJ...478..358D}
}

@ARTICLE{2022MNRAS.514.4932C,
       author = {{Chahal}, Deepak and {de Grijs}, Richard and {Kamath}, Devika and {Chen}, Xiaodian},
        title = "{Statistics of BY Draconis chromospheric variable stars}",
      journal = {\mnras},
         year = 2022,
        month = aug,
       volume = {514},
       number = {4},
        pages = {4932-4943},
          doi = {10.1093/mnras/stac1660},
archivePrefix = {arXiv},
       eprint = {2206.05505},
 primaryClass = {astro-ph.SR},
       adsurl = {https://ui.adsabs.harvard.edu/abs/2022MNRAS.514.4932C}
}

@ARTICLE{1973MNRAS.164..343B,
       author = {{Bopp}, B.~W. and {Evans}, D.~S.},
        title = "{The spotted flare stars BY Dra and CC Eri: a model for the spots and some astrophysical implications.}",
      journal = {\mnras},
         year = 1973,
        month = jan,
       volume = {164},
        pages = {343-356},
          doi = {10.1093/mnras/164.4.343},
       adsurl = {https://ui.adsabs.harvard.edu/abs/1973MNRAS.164..343B}
}

@ARTICLE{2024A&A...689A.103Z,
       author = {{Zhang}, Liyun and {Yang}, Zilu and {Su}, Tianhao and {Han}, Xianming L. and {Misra}, Prabhakar},
        title = "{Properties of flare events based on light curves from the TESS survey: II. 20-second cadence}",
      journal = {\aap},
         year = 2024,
        month = sep,
       volume = {689},
          eid = {A103},
        pages = {A103},
          doi = {10.1051/0004-6361/202348343},
       adsurl = {https://ui.adsabs.harvard.edu/abs/2024A&A...689A.103Z}
}

@ARTICLE{2025A&A...694A.161S,
       author = {{Seli}, B. and {Vida}, K. and {Ol{\'a}h}, K. and {G{\"o}rgei}, A. and {So{\'o}s}, Sz. and {P{\'a}l}, A. and {Kriskovics}, L. and {K{\H{o}}v{\'a}ri}, Zs.},
        title = "{Stellar flare morphology with TESS across the main sequence}",
      journal = {\aap},
         year = 2025,
        month = feb,
       volume = {694},
          eid = {A161},
        pages = {A161},
          doi = {10.1051/0004-6361/202452489},
archivePrefix = {arXiv},
       eprint = {2412.12989},
 primaryClass = {astro-ph.SR},
       adsurl = {https://ui.adsabs.harvard.edu/abs/2025A&A...694A.161S}
}

@ARTICLE{2023ApJ...943..110S,
       author = {{Sahai}, Raghvendra and {Bujarrabal}, Valentin and {Quintana-Lacaci}, Guillermo and {Reindl}, Nicole and {Van de Steene}, Griet and {S{\'a}nchez Contreras}, Carmen and {Ressler}, Michael E.},
        title = "{The Binary and the Disk: The Beauty is Found within NGC3132 with JWST}",
      journal = {\apj},
         year = 2023,
        month = feb,
       volume = {943},
       number = {2},
          eid = {110},
        pages = {110},
          doi = {10.3847/1538-4357/aca7ba},
archivePrefix = {arXiv},
       eprint = {2211.16741},
 primaryClass = {astro-ph.SR},
       adsurl = {https://ui.adsabs.harvard.edu/abs/2023ApJ...943..110S}
}

@ARTICLE{2005ApJ...619L...1M,
       author = {{Martin}, D. Christopher and {Fanson}, James and {Schiminovich}, David and {Morrissey}, Patrick and {Friedman}, Peter G. and {Barlow}, Tom A. and {Conrow}, Tim and {Grange}, Robert and {Jelinsky}, Patrick N. and {Milliard}, Bruno and {Siegmund}, Oswald H.~W. and {Bianchi}, Luciana and {Byun}, Yong-Ik and {Donas}, Jose and {Forster}, Karl and {Heckman}, Timothy M. and {Lee}, Young-Wook and {Madore}, Barry F. and {Malina}, Roger F. and {Neff}, Susan G. and {Rich}, R. Michael and {Small}, Todd and {Surber}, Frank and {Szalay}, Alex S. and {Welsh}, Barry and {Wyder}, Ted K.},
        title = "{The Galaxy Evolution Explorer: A Space Ultraviolet Survey Mission}",
      journal = {\apjl},
         year = 2005,
        month = jan,
       volume = {619},
       number = {1},
        pages = {L1-L6},
          doi = {10.1086/426387},
archivePrefix = {arXiv},
       eprint = {astro-ph/0411302},
 primaryClass = {astro-ph},
       adsurl = {https://ui.adsabs.harvard.edu/abs/2005ApJ...619L...1M}
}

@ARTICLE{2023A&A...669A..45T,
       author = {{Temmink}, K.~D. and {Pols}, O.~R. and {Justham}, S. and {Istrate}, A.~G. and {Toonen}, S.},
        title = "{Coping with loss. Stability of mass transfer from post-main-sequence donor stars}",
      journal = {\aap},
         year = 2023,
        month = jan,
       volume = {669},
          eid = {A45},
        pages = {A45},
          doi = {10.1051/0004-6361/202244137},
archivePrefix = {arXiv},
       eprint = {2209.12707},
 primaryClass = {astro-ph.SR},
       adsurl = {https://ui.adsabs.harvard.edu/abs/2023A&A...669A..45T}
}

@ARTICLE{2025PASP..137j4205Y,
       author = {{Yamaguchi}, Natsuko and {El-Badry}, Kareem and {Shahaf}, Sahar},
        title = "{Population Demographics of White Dwarf Binaries with Intermediate Separations: Gaia Constraints on post-AGB Mass Transfer}",
      journal = {\pasp},
         year = 2025,
        month = oct,
       volume = {137},
       number = {10},
          eid = {104205},
        pages = {104205},
          doi = {10.1088/1538-3873/ae0d30},
archivePrefix = {arXiv},
       eprint = {2505.14786},
 primaryClass = {astro-ph.SR},
       adsurl = {https://ui.adsabs.harvard.edu/abs/2025PASP..137j4205Y}
}

@ARTICLE{2025ApJ...991..226H,
       author = {{Hwang}, Hsiang-Chih and {Zakamska}, Nadia L.},
        title = "{White Dwarfs in Wide Binaries: The Strong Effects of Stellar Evolution and Mass Loss}",
      journal = {\apj},
         year = 2025,
        month = oct,
       volume = {991},
       number = {2},
          eid = {226},
        pages = {226},
          doi = {10.3847/1538-4357/adfa1c},
archivePrefix = {arXiv},
       eprint = {2508.08364},
 primaryClass = {astro-ph.SR},
       adsurl = {https://ui.adsabs.harvard.edu/abs/2025ApJ...991..226H}
}

@ARTICLE{2023ApJ...955L..14S,
       author = {{Shariat}, Cheyanne and {Naoz}, Smadar and {Hansen}, Bradley M.~S. and {Angelo}, Isabel and {Michaely}, Erez and {Stephan}, Alexander P.},
        title = "{Dynamical Evolution of White Dwarfs in Triples in the Era of Gaia}",
      journal = {\apjl},
         year = 2023,
        month = sep,
       volume = {955},
       number = {1},
          eid = {L14},
        pages = {L14},
          doi = {10.3847/2041-8213/acf76b},
archivePrefix = {arXiv},
       eprint = {2306.13130},
 primaryClass = {astro-ph.SR},
       adsurl = {https://ui.adsabs.harvard.edu/abs/2023ApJ...955L..14S}
}

@ARTICLE{2016ComAC...3....6T,
       author = {{Toonen}, Silvia and {Hamers}, Adrian and {Portegies Zwart}, Simon},
        title = "{The evolution of hierarchical triple star-systems}",
      journal = {Computational Astrophysics and Cosmology},
         year = 2016,
        month = dec,
       volume = {3},
       number = {1},
          eid = {6},
        pages = {6},
          doi = {10.1186/s40668-016-0019-0},
archivePrefix = {arXiv},
       eprint = {1612.06172},
 primaryClass = {astro-ph.SR},
       adsurl = {https://ui.adsabs.harvard.edu/abs/2016ComAC...3....6T}
}

@ARTICLE{2020A&A...640A..16T,
       author = {{Toonen}, S. and {Portegies Zwart}, S. and {Hamers}, A.~S. and {Bandopadhyay}, D.},
        title = "{The evolution of stellar triples. The most common evolutionary pathways}",
      journal = {\aap},
         year = 2020,
        month = aug,
       volume = {640},
          eid = {A16},
        pages = {A16},
          doi = {10.1051/0004-6361/201936835},
archivePrefix = {arXiv},
       eprint = {2004.07848},
 primaryClass = {astro-ph.SR},
       adsurl = {https://ui.adsabs.harvard.edu/abs/2020A&A...640A..16T}
}

@ARTICLE{2022A&A...661A..61T,
       author = {{Toonen}, S. and {Boekholt}, T.~C.~N. and {Portegies Zwart}, S.},
        title = "{Stellar triples on the edge. Comprehensive overview of the evolution of destabilised triples leading to stellar and binary exotica}",
      journal = {\aap},
         year = 2022,
        month = may,
       volume = {661},
          eid = {A61},
        pages = {A61},
          doi = {10.1051/0004-6361/202141991},
archivePrefix = {arXiv},
       eprint = {2108.04272},
 primaryClass = {astro-ph.SR},
       adsurl = {https://ui.adsabs.harvard.edu/abs/2022A&A...661A..61T}
}

@ARTICLE{2022MNRAS.514.1895K,
       author = {{Knigge}, C. and {Toonen}, S. and {Boekholt}, T.~C.~N.},
        title = "{A triple star origin for T Pyx and other short-period recurrent novae}",
      journal = {\mnras},
         year = 2022,
        month = aug,
       volume = {514},
       number = {2},
        pages = {1895-1907},
          doi = {10.1093/mnras/stac1336},
archivePrefix = {arXiv},
       eprint = {2205.00014},
 primaryClass = {astro-ph.SR},
       adsurl = {https://ui.adsabs.harvard.edu/abs/2022MNRAS.514.1895K}
}

@ARTICLE{2023ApJ...950....9R,
       author = {{Rajamuthukumar}, Abinaya Swaruba and {Hamers}, Adrian S. and {Neunteufel}, Patrick and {Pakmor}, R{\"u}diger and {de Mink}, Selma E.},
        title = "{Triple Evolution: An Important Channel in the Formation of Type Ia Supernovae}",
      journal = {\apj},
         year = 2023,
        month = jun,
       volume = {950},
       number = {1},
          eid = {9},
        pages = {9},
          doi = {10.3847/1538-4357/acc86c},
archivePrefix = {arXiv},
       eprint = {2211.04463},
 primaryClass = {astro-ph.SR},
       adsurl = {https://ui.adsabs.harvard.edu/abs/2023ApJ...950....9R}
}

@ARTICLE{2025A&A...704A.156R,
       author = {{Rajamuthukumar}, Abinaya Swaruba and {Korol}, Valeriya and {Stegmann}, Jakob and {Preece}, Holly and {Pakmor}, R{\"u}diger and {Justham}, Stephen and {Toonen}, Silvia and {de Mink}, Selma E.},
        title = "{The role of triple evolution in the formation of LISA double white dwarfs}",
      journal = {\aap},
         year = 2025,
        month = dec,
       volume = {704},
          eid = {A156},
        pages = {A156},
          doi = {10.1051/0004-6361/202554277},
archivePrefix = {arXiv},
       eprint = {2502.09607},
 primaryClass = {astro-ph.SR},
       adsurl = {https://ui.adsabs.harvard.edu/abs/2025A&A...704A.156R}
}

@ARTICLE{2025ApJ...978...47S,
       author = {{Shariat}, Cheyanne and {Naoz}, Smadar and {El-Badry}, Kareem and {Rodriguez}, Antonio C. and {Hansen}, Bradley M.~S. and {Angelo}, Isabel and {Stephan}, Alexander P.},
        title = "{Once a Triple, Not Always a Triple: The Evolution of Hierarchical Triples That Yield Merged Inner Binaries}",
      journal = {\apj},
         year = 2025,
        month = jan,
       volume = {978},
       number = {1},
          eid = {47},
        pages = {47},
          doi = {10.3847/1538-4357/ad944a},
archivePrefix = {arXiv},
       eprint = {2407.06257},
 primaryClass = {astro-ph.SR},
       adsurl = {https://ui.adsabs.harvard.edu/abs/2025ApJ...978...47S}
}

@INPROCEEDINGS{2026enap....2..279P,
       author = {{Perets}, Hagai B.},
        title = "{Evolution of triple stars}",
    booktitle = {Encyclopedia of Astrophysics, Volume 2},
         year = 2026,
       volume = {2},
        month = jan,
        pages = {279-297},
          doi = {10.1016/B978-0-443-21439-4.00108-5},
archivePrefix = {arXiv},
       eprint = {2504.02939},
 primaryClass = {astro-ph.SR},
       adsurl = {https://ui.adsabs.harvard.edu/abs/2026enap....2..279P}
}

@ARTICLE{2024ApJ...970L..11H,
       author = {{Hallakoun}, Na'ama and {Shahaf}, Sahar and {Mazeh}, Tsevi and {Toonen}, Silvia and {Ben-Ami}, Sagi},
        title = "{A Deficit of Massive White Dwarfs in Gaia Astrometric Binaries}",
      journal = {\apjl},
         year = 2024,
        month = jul,
       volume = {970},
       number = {1},
          eid = {L11},
        pages = {L11},
          doi = {10.3847/2041-8213/ad5e63},
archivePrefix = {arXiv},
       eprint = {2311.17145},
 primaryClass = {astro-ph.SR},
       adsurl = {https://ui.adsabs.harvard.edu/abs/2024ApJ...970L..11H}
}

@ARTICLE{1999A&A...349..389V,
       author = {{Voges}, W. and {Aschenbach}, B. and {Boller}, Th. and {Br{\"a}uninger}, H. and {Briel}, U. and {Burkert}, W. and {Dennerl}, K. and {Englhauser}, J. and {Gruber}, R. and {Haberl}, F. and {Hartner}, G. and {Hasinger}, G. and {K{\"u}rster}, M. and {Pfeffermann}, E. and {Pietsch}, W. and {Predehl}, P. and {Rosso}, C. and {Schmitt}, J.~H.~M.~M. and {Tr{\"u}mper}, J. and {Zimmermann}, H.~U.},
        title = "{The ROSAT all-sky survey bright source catalogue}",
      journal = {\aap},
         year = 1999,
        month = sep,
       volume = {349},
        pages = {389-405},
          doi = {10.48550/arXiv.astro-ph/9909315},
archivePrefix = {arXiv},
       eprint = {astro-ph/9909315},
 primaryClass = {astro-ph},
       adsurl = {https://ui.adsabs.harvard.edu/abs/1999A&A...349..389V}
}

@ARTICLE{2022ApJ...924...31B,
       author = {{Basri}, Gibor and {Streichenberger}, Tristan and {McWard}, Connor and {Edmond}, IV, Lawrence and {Tan}, Joanne and {Lee}, Minjoo and {Melton}, Trey},
        title = "{A New Method for Estimating Starspot Lifetimes Based on Autocorrelation Functions}",
      journal = {\apj},
         year = 2022,
        month = jan,
       volume = {924},
       number = {1},
          eid = {31},
        pages = {31},
          doi = {10.3847/1538-4357/ac3420},
archivePrefix = {arXiv},
       eprint = {2110.13284},
 primaryClass = {astro-ph.SR},
       adsurl = {https://ui.adsabs.harvard.edu/abs/2022ApJ...924...31B}
}

@ARTICLE{2024MNRAS.528.3392W,
       author = {{Wesson}, R. and {Matsuura}, Mikako and {Zijlstra}, Albert A. and {Volk}, Kevin and {Kavanagh}, Patrick J. and {Garc{\'\i}a-Segura}, Guillermo and {McDonald}, I. and {Sahai}, Raghvendra and {Barlow}, M.~J. and {Cox}, Nick L.~J. and {Bernard-Salas}, Jeronimo and {Aleman}, Isabel and {Cami}, Jan and {Clark}, Nicholas and {Dinerstein}, Harriet L. and {Justtanont}, K. and {Kaplan}, Kyle F. and {Manchado}, A. and {Peeters}, Els and {Van de Steene}, Griet C. and {van Hoof}, Peter A.~M.},
        title = "{JWST observations of the Ring Nebula (NGC 6720): I. Imaging of the rings, globules, and arcs}",
      journal = {\mnras},
         year = 2024,
        month = feb,
       volume = {528},
       number = {2},
        pages = {3392-3416},
          doi = {10.1093/mnras/stad3670},
archivePrefix = {arXiv},
       eprint = {2308.09027},
 primaryClass = {astro-ph.SR},
       adsurl = {https://ui.adsabs.harvard.edu/abs/2024MNRAS.528.3392W}
}

@ARTICLE{2007ApJS..173..682M,
       author = {{Morrissey}, Patrick and {Conrow}, Tim and {Barlow}, Tom A. and {Small}, Todd and {Seibert}, Mark and {Wyder}, Ted K. and {Budav{\'a}ri}, Tam{\'a}s and {Arnouts}, Stephane and {Friedman}, Peter G. and {Forster}, Karl and {Martin}, D. Christopher and {Neff}, Susan G. and {Schiminovich}, David and {Bianchi}, Luciana and {Donas}, Jos{\'e} and {Heckman}, Timothy M. and {Lee}, Young-Wook and {Madore}, Barry F. and {Milliard}, Bruno and {Rich}, R. Michael and {Szalay}, Alex S. and {Welsh}, Barry Y. and {Yi}, Sukyoung K.},
        title = "{The Calibration and Data Products of GALEX}",
      journal = {\apjs},
         year = 2007,
        month = dec,
       volume = {173},
       number = {2},
        pages = {682-697},
          doi = {10.1086/520512},
archivePrefix = {arXiv},
       eprint = {0706.0755},
 primaryClass = {astro-ph},
       adsurl = {https://ui.adsabs.harvard.edu/abs/2007ApJS..173..682M}
}

@ARTICLE{2022MNRAS.509.2566M,
       author = {{Molnar}, Thomas A. and {Sanders}, Jason L. and {Smith}, Leigh C. and {Belokurov}, Vasily and {Lucas}, Philip and {Minniti}, Dante},
        title = "{Variable star classification across the Galactic bulge and disc with the VISTA Variables in the V{\'\i}a L{\'a}ctea survey}",
      journal = {\mnras},
         year = 2022,
        month = jan,
       volume = {509},
       number = {2},
        pages = {2566-2592},
          doi = {10.1093/mnras/stab3116},
archivePrefix = {arXiv},
       eprint = {2110.15371},
 primaryClass = {astro-ph.SR},
       adsurl = {https://ui.adsabs.harvard.edu/abs/2022MNRAS.509.2566M}
}

@ARTICLE{2025ApJ...980..227C,
       author = {{Chen}, Pinjian and {Fang}, Xuan and {Chen}, Xiaodian and {Liu}, Jifeng},
        title = "{Periodic Variability of the Central Stars of Planetary Nebulae Surveyed through the Zwicky Transient Facility}",
      journal = {\apj},
         year = 2025,
        month = feb,
       volume = {980},
       number = {2},
          eid = {227},
        pages = {227},
          doi = {10.3847/1538-4357/ada94a},
archivePrefix = {arXiv},
       eprint = {2501.06056},
 primaryClass = {astro-ph.SR},
       adsurl = {https://ui.adsabs.harvard.edu/abs/2025ApJ...980..227C}
}

@ARTICLE{2017A&A...600L...9J,
       author = {{Jones}, D. and {Van Winckel}, H. and {Aller}, A. and {Exter}, K. and {De Marco}, O.},
        title = "{The long-period binary central stars of the planetary nebulae NGC 1514 and LoTr 5}",
      journal = {\aap},
         year = 2017,
        month = apr,
       volume = {600},
          eid = {L9},
        pages = {L9},
          doi = {10.1051/0004-6361/201730700},
archivePrefix = {arXiv},
       eprint = {1703.05096},
 primaryClass = {astro-ph.SR},
       adsurl = {https://ui.adsabs.harvard.edu/abs/2017A&A...600L...9J}
}

@ARTICLE{2018ApJ...860L..17E,
       author = {{El-Badry}, Kareem and {Rix}, Hans-Walter and {Weisz}, Daniel R.},
        title = "{An Empirical Measurement of the Initial-Final Mass Relation with Gaia White Dwarfs}",
      journal = {\apjl},
         year = 2018,
        month = jun,
       volume = {860},
       number = {2},
          eid = {L17},
        pages = {L17},
          doi = {10.3847/2041-8213/aaca9c},
archivePrefix = {arXiv},
       eprint = {1805.05849},
 primaryClass = {astro-ph.SR},
       adsurl = {https://ui.adsabs.harvard.edu/abs/2018ApJ...860L..17E}
}

@ARTICLE{2024MNRAS.527.3602C,
       author = {{Cunningham}, Tim and {Tremblay}, Pier-Emmanuel and {W. O'Brien}, Mairi},
        title = "{Initial-final mass relation from white dwarfs within 40 pc}",
      journal = {\mnras},
         year = 2024,
        month = jan,
       volume = {527},
       number = {2},
        pages = {3602-3611},
          doi = {10.1093/mnras/stad3275},
archivePrefix = {arXiv},
       eprint = {2310.15410},
 primaryClass = {astro-ph.SR},
       adsurl = {https://ui.adsabs.harvard.edu/abs/2024MNRAS.527.3602C}
}

@ARTICLE{2025OJAp....8E..62E,
       author = {{El-Badry}, Kareem},
        title = "{How to use Gaia parallaxes for stars with poor astrometric fits}",
      journal = {The Open Journal of Astrophysics},
         year = 2025,
        month = may,
       volume = {8},
          eid = {62},
        pages = {62},
          doi = {10.33232/001c.138448},
archivePrefix = {arXiv},
       eprint = {2504.11528},
 primaryClass = {astro-ph.SR},
       adsurl = {https://ui.adsabs.harvard.edu/abs/2025OJAp....8E..62E}
}

@ARTICLE{2020ApJ...899..132G,
       author = {{Ge}, Hongwei and {Webbink}, Ronald F. and {Chen}, Xuefei and {Han}, Zhanwen},
        title = "{Adiabatic Mass Loss in Binary Stars. III. From the Base of the Red Giant Branch to the Tip of the Asymptotic Giant Branch}",
      journal = {\apj},
         year = 2020,
        month = aug,
       volume = {899},
       number = {2},
          eid = {132},
        pages = {132},
          doi = {10.3847/1538-4357/aba7b7},
archivePrefix = {arXiv},
       eprint = {2007.09848},
 primaryClass = {astro-ph.SR},
       adsurl = {https://ui.adsabs.harvard.edu/abs/2020ApJ...899..132G}
}

@ARTICLE{2016A&A...595A...1G,
       author = {{Gaia Collaboration} and {Prusti}, T. and {de Bruijne}, J.~H.~J. and {Brown}, A.~G.~A. and {Vallenari}, A. and {Babusiaux}, C. and {Bailer-Jones}, C.~A.~L. and {Bastian}, U. and {Biermann}, M. and {Evans}, D.~W. and {Eyer}, L. and {Jansen}, F. and {Jordi}, C. and {Klioner}, S.~A. and {Lammers}, U. and {Lindegren}, L. and {Luri}, X. and {Mignard}, F. and {Milligan}, D.~J. and {Panem}, C. and {Poinsignon}, V. and {Pourbaix}, D. and {Randich}, S. and {Sarri}, G. and {Sartoretti}, P. and {Siddiqui}, H.~I. and {Soubiran}, C. and {Valette}, V. and {van Leeuwen}, F. and {Walton}, N.~A. and {Aerts}, C. and {Arenou}, F. and {Cropper}, M. and {Drimmel}, R. and {H{\o}g}, E. and {Katz}, D. and {Lattanzi}, M.~G. and {O'Mullane}, W. and {Grebel}, E.~K. and {Holland}, A.~D. and {Huc}, C. and {Passot}, X. and {Bramante}, L. and {Cacciari}, C. and {Casta{\~n}eda}, J. and {Chaoul}, L. and {Cheek}, N. and {De Angeli}, F. and {Fabricius}, C. and {Guerra}, R. and {Hern{\'a}ndez}, J. and {Jean-Antoine-Piccolo}, A. and {Masana}, E. and {Messineo}, R. and {Mowlavi}, N. and {Nienartowicz}, K. and {Ord{\'o}{\~n}ez-Blanco}, D. and {Panuzzo}, P. and {Portell}, J. and {Richards}, P.~J. and {Riello}, M. and {Seabroke}, G.~M. and {Tanga}, P. and {Th{\'e}venin}, F. and {Torra}, J. and {Els}, S.~G. and {Gracia-Abril}, G. and {Comoretto}, G. and {Garcia-Reinaldos}, M. and {Lock}, T. and {Mercier}, E. and {Altmann}, M. and {Andrae}, R. and {Astraatmadja}, T.~L. and {Bellas-Velidis}, I. and {Benson}, K. and {Berthier}, J. and {Blomme}, R. and {Busso}, G. and {Carry}, B. and {Cellino}, A. and {Clementini}, G. and {Cowell}, S. and {Creevey}, O. and {Cuypers}, J. and {Davidson}, M. and {De Ridder}, J. and {de Torres}, A. and {Delchambre}, L. and {Dell'Oro}, A. and {Ducourant}, C. and {Fr{\'e}mat}, Y. and {Garc{\'\i}a-Torres}, M. and {Gosset}, E. and {Halbwachs}, J.-L. and {Hambly}, N.~C. and {Harrison}, D.~L. and {Hauser}, M. and {Hestroffer}, D. and {Hodgkin}, S.~T. and {Huckle}, H.~E. and {Hutton}, A. and {Jasniewicz}, G. and {Jordan}, S. and {Kontizas}, M. and {Korn}, A.~J. and {Lanzafame}, A.~C. and {Manteiga}, M. and {Moitinho}, A. and {Muinonen}, K. and {Osinde}, J. and {Pancino}, E. and {Pauwels}, T. and {Petit}, J.-M. and {Recio-Blanco}, A. and {Robin}, A.~C. and {Sarro}, L.~M. and {Siopis}, C. and {Smith}, M. and {Smith}, K.~W. and {Sozzetti}, A. and {Thuillot}, W. and {van Reeven}, W. and {Viala}, Y. and {Abbas}, U. and {Abreu Aramburu}, A. and {Accart}, S. and {Aguado}, J.~J. and {Allan}, P.~M. and {Allasia}, W. and {Altavilla}, G. and {{\'A}lvarez}, M.~A. and {Alves}, J. and {Anderson}, R.~I. and {Andrei}, A.~H. and {Anglada Varela}, E. and {Antiche}, E. and {Antoja}, T. and {Ant{\'o}n}, S. and {Arcay}, B. and {Atzei}, A. and {Ayache}, L. and {Bach}, N. and {Baker}, S.~G. and {Balaguer-N{\'u}{\~n}ez}, L. and {Barache}, C. and {Barata}, C. and {Barbier}, A. and {Barblan}, F. and {Baroni}, M. and {Barrado y Navascu{\'e}s}, D. and {Barros}, M. and {Barstow}, M.~A. and {Becciani}, U. and {Bellazzini}, M. and {Bellei}, G. and {Bello Garc{\'\i}a}, A. and {Belokurov}, V. and {Bendjoya}, P. and {Berihuete}, A. and {Bianchi}, L. and {Bienaym{\'e}}, O. and {Billebaud}, F. and {Blagorodnova}, N. and {Blanco-Cuaresma}, S. and {Boch}, T. and {Bombrun}, A. and {Borrachero}, R. and {Bouquillon}, S. and {Bourda}, G. and {Bouy}, H. and {Bragaglia}, A. and {Breddels}, M.~A. and {Brouillet}, N. and {Br{\"u}semeister}, T. and {Bucciarelli}, B. and {Budnik}, F. and {Burgess}, P. and {Burgon}, R. and {Burlacu}, A. and {Busonero}, D. and {Buzzi}, R. and {Caffau}, E. and {Cambras}, J. and {Campbell}, H. and {Cancelliere}, R. and {Cantat-Gaudin}, T. and {Carlucci}, T. and {Carrasco}, J.~M. and {Castellani}, M. and {Charlot}, P. and {Charnas}, J. and {Charvet}, P. and {Chassat}, F. and {Chiavassa}, A. and {Clotet}, M. and {Cocozza}, G. and {Collins}, R.~S. and {Collins}, P. and {Costigan}, G.},
        title = "{The Gaia mission}",
      journal = {\aap},
         year = 2016,
        month = nov,
       volume = {595},
          eid = {A1},
        pages = {A1},
          doi = {10.1051/0004-6361/201629272},
archivePrefix = {arXiv},
       eprint = {1609.04153},
 primaryClass = {astro-ph.IM},
       adsurl = {https://ui.adsabs.harvard.edu/abs/2016A&A...595A...1G}
}

@ARTICLE{2018MNRAS.480.4884E,
       author = {{El-Badry}, Kareem and {Rix}, Hans-Walter},
        title = "{Imprints of white dwarf recoil in the separation distribution of Gaia wide binaries}",
      journal = {\mnras},
         year = 2018,
        month = nov,
       volume = {480},
       number = {4},
        pages = {4884-4902},
          doi = {10.1093/mnras/sty2186},
archivePrefix = {arXiv},
       eprint = {1807.06011},
 primaryClass = {astro-ph.SR},
       adsurl = {https://ui.adsabs.harvard.edu/abs/2018MNRAS.480.4884E}
}

@ARTICLE{1996A&A...316..147F,
       author = {{Fleming}, T.~A. and {Snowden}, S.~L. and {Pfeffermann}, E. and {Briel}, U. and {Greiner}, J.},
        title = "{Catalogue and luminosity function of white dwarfs detected in the ROSAT all-sky survey.}",
      journal = {\aap},
         year = 1996,
        month = dec,
       volume = {316},
        pages = {147-154},
       adsurl = {https://ui.adsabs.harvard.edu/abs/1996A&A...316..147F}
}

@ARTICLE{1993MNRAS.264...16B,
       author = {{Barstow}, M.~A. and {Fleming}, T.~A. and {Diamond}, C.~J. and {Finley}, D.~S. and {Sansom}, A.~E. and {Rosen}, S.~R. and {Koester}, D. and {Marsh}, M.~C. and {Holberg}, J.~B. and {Kidder}, K.},
        title = "{ROSAT studies of the composition and structure of DA white dwarf atmospheres.}",
      journal = {\mnras},
         year = 1993,
        month = sep,
       volume = {264},
        pages = {16},
          doi = {10.1093/mnras/264.1.16},
       adsurl = {https://ui.adsabs.harvard.edu/abs/1993MNRAS.264...16B}
}

@ARTICLE{1994A&A...282..586M,
       author = {{Murset}, Urs and {Nussbaumer}, Harry},
        title = "{Temperatures and luminosities of symbiotic novae.}",
      journal = {\aap},
         year = 1994,
        month = feb,
       volume = {282},
        pages = {586-604},
       adsurl = {https://ui.adsabs.harvard.edu/abs/1994A&A...282..586M}
}

@ARTICLE{2010PASA...27..220W,
       author = {{Wareing}, C.~J.},
        title = "{Rebrightening of Planetary Nebulae Through Interaction with the Interstellar Medium}",
      journal = {\pasa},
         year = 2010,
        month = may,
       volume = {27},
       number = {2},
        pages = {220-226},
          doi = {10.1071/AS09035},
archivePrefix = {arXiv},
       eprint = {0910.2200},
 primaryClass = {astro-ph.SR},
       adsurl = {https://ui.adsabs.harvard.edu/abs/2010PASA...27..220W}
}

@ARTICLE{2006MNRAS.366..387W,
       author = {{Wareing}, C.~J. and {O'Brien}, T.~J. and {Zijlstra}, Albert A. and {Kwitter}, K.~B. and {Irwin}, J. and {Wright}, N. and {Greimel}, R. and {Drew}, J.~E.},
        title = "{The shaping of planetary nebula Sh2-188 through interactionwith the interstellar medium}",
      journal = {\mnras},
         year = 2006,
        month = feb,
       volume = {366},
       number = {2},
        pages = {387-396},
          doi = {10.1111/j.1365-2966.2005.09875.x},
archivePrefix = {arXiv},
       eprint = {astro-ph/0512028},
 primaryClass = {astro-ph},
       adsurl = {https://ui.adsabs.harvard.edu/abs/2006MNRAS.366..387W}
}

@ARTICLE{1990ApJ...360..173B,
       author = {{Borkowski}, Kazimierz J. and {Sarazin}, Craig L. and {Soker}, Noam},
        title = "{Interaction of Planetary Nebulae with the Interstellar Medium}",
      journal = {\apj},
         year = 1990,
        month = sep,
       volume = {360},
        pages = {173},
          doi = {10.1086/169106},
       adsurl = {https://ui.adsabs.harvard.edu/abs/1990ApJ...360..173B}
}

@ARTICLE{2016MNRAS.457....9C,
       author = {{Chiotellis}, A. and {Boumis}, P. and {Nanouris}, N. and {Meaburn}, J. and {Dimitriadis}, G.},
        title = "{Modelling the cometary structure of the planetary nebula HFG1 based on the evolution of its binary central star V664 Cas}",
      journal = {\mnras},
         year = 2016,
        month = mar,
       volume = {457},
       number = {1},
        pages = {9-23},
          doi = {10.1093/mnras/stv2798},
archivePrefix = {arXiv},
       eprint = {1512.00864},
 primaryClass = {astro-ph.SR},
       adsurl = {https://ui.adsabs.harvard.edu/abs/2016MNRAS.457....9C}
}

@ARTICLE{2023A&A...680A..99M,
       author = {{Martinez}, J.~R. and {del Palacio}, S. and {Bosch-Ramon}, V.},
        title = "{Probing the non-thermal physics of stellar bow shocks using radio observations}",
      journal = {\aap},
         year = 2023,
        month = dec,
       volume = {680},
          eid = {A99},
        pages = {A99},
          doi = {10.1051/0004-6361/202347720},
archivePrefix = {arXiv},
       eprint = {2310.18669},
 primaryClass = {astro-ph.HE},
       adsurl = {https://ui.adsabs.harvard.edu/abs/2023A&A...680A..99M}
}

@ARTICLE{2021MNRAS.506.2269E,
       author = {{El-Badry}, Kareem and {Rix}, Hans-Walter and {Heintz}, Tyler M.},
        title = "{A million binaries from Gaia eDR3: sample selection and validation of Gaia parallax uncertainties}",
      journal = {\mnras},
         year = 2021,
        month = sep,
       volume = {506},
       number = {2},
        pages = {2269-2295},
          doi = {10.1093/mnras/stab323},
archivePrefix = {arXiv},
       eprint = {2101.05282},
 primaryClass = {astro-ph.SR},
       adsurl = {https://ui.adsabs.harvard.edu/abs/2021MNRAS.506.2269E}
}

@ARTICLE{2020RNAAS...4..204H,
       author = {{Huang}, Chelsea X. and {Vanderburg}, Andrew and {P{\'a}l}, Andras and {Sha}, Lizhou and {Yu}, Liang and {Fong}, Willie and {Fausnaugh}, Michael and {Shporer}, Avi and {Guerrero}, Natalia and {Vanderspek}, Roland and {Ricker}, George},
        title = "{Photometry of 10 Million Stars from the First Two Years of TESS Full Frame Images: Part I}",
      journal = {Research Notes of the American Astronomical Society},
         year = 2020,
        month = nov,
       volume = {4},
       number = {11},
          eid = {204},
        pages = {204},
          doi = {10.3847/2515-5172/abca2e},
archivePrefix = {arXiv},
       eprint = {2011.06459},
 primaryClass = {astro-ph.EP},
       adsurl = {https://ui.adsabs.harvard.edu/abs/2020RNAAS...4..204H}
}

@ARTICLE{2009MNRAS.396.1186B,
       author = {{Boumis}, P. and {Meaburn}, J. and {Lloyd}, M. and {Akras}, S.},
        title = "{A long trail behind the planetary nebula HFG1 (PK 136+05) and its pre-cataclysmic binary central star V664 Cas}",
      journal = {\mnras},
         year = 2009,
        month = jun,
       volume = {396},
       number = {2},
        pages = {1186-1188},
          doi = {10.1111/j.1365-2966.2009.14784.x},
archivePrefix = {arXiv},
       eprint = {0903.2852},
 primaryClass = {astro-ph.GA},
       adsurl = {https://ui.adsabs.harvard.edu/abs/2009MNRAS.396.1186B}
}

@ARTICLE{2007MNRAS.382.1233W,
       author = {{Wareing}, C.~J. and {Zijlstra}, Albert A. and {O'Brien}, T.~J.},
        title = "{The interaction of planetary nebulae and their asymptotic giant branch progenitors with the interstellar medium}",
      journal = {\mnras},
         year = 2007,
        month = nov,
       volume = {382},
       number = {3},
        pages = {1233-1245},
          doi = {10.1111/j.1365-2966.2007.12459.x},
archivePrefix = {arXiv},
       eprint = {0709.2848},
 primaryClass = {astro-ph},
       adsurl = {https://ui.adsabs.harvard.edu/abs/2007MNRAS.382.1233W}
}

@ARTICLE{2025A&A...696A.243G,
       author = {{Godoy-Rivera}, D. and {Mathur}, S. and {Garc{\'\i}a}, R.~A. and {Pinsonneault}, M.~H. and {Santos}, {\^A}. R.~G. and {Beck}, P.~G. and {Grossmann}, D.~H. and {Schimak}, L. and {Bedell}, M. and {Merc}, J. and {Escorza}, A.},
        title = "{Kepler meets Gaia DR3: Homogeneous extinction-corrected color-magnitude diagram and binary classification}",
      journal = {\aap},
         year = 2025,
        month = apr,
       volume = {696},
          eid = {A243},
        pages = {A243},
          doi = {10.1051/0004-6361/202348735},
archivePrefix = {arXiv},
       eprint = {2501.18719},
 primaryClass = {astro-ph.SR},
       adsurl = {https://ui.adsabs.harvard.edu/abs/2025A&A...696A.243G}
}

@ARTICLE{2022ApJS..258...16P,
       author = {{Pr{\v{s}}a}, Andrej and {Kochoska}, Angela and {Conroy}, Kyle E. and {Eisner}, Nora and {Hey}, Daniel R. and {IJspeert}, Luc and {Kruse}, Ethan and {Fleming}, Scott W. and {Johnston}, Cole and {Kristiansen}, Martti H. and {LaCourse}, Daryll and {Mortensen}, Danielle and {Pepper}, Joshua and {Stassun}, Keivan G. and {Torres}, Guillermo and {Abdul-Masih}, Michael and {Chakraborty}, Joheen and {Gagliano}, Robert and {Guo}, Zhao and {Hambleton}, Kelly and {Hong}, Kyeongsoo and {Jacobs}, Thomas and {Jones}, David and {Kostov}, Veselin and {Lee}, Jae Woo and {Omohundro}, Mark and {Orosz}, Jerome A. and {Page}, Emma J. and {Powell}, Brian P. and {Rappaport}, Saul and {Reed}, Phill and {Schnittman}, Jeremy and {Schwengeler}, Hans Martin and {Shporer}, Avi and {Terentev}, Ivan A. and {Vanderburg}, Andrew and {Welsh}, William F. and {Caldwell}, Douglas A. and {Doty}, John P. and {Jenkins}, Jon M. and {Latham}, David W. and {Ricker}, George R. and {Seager}, Sara and {Schlieder}, Joshua E. and {Shiao}, Bernie and {Vanderspek}, Roland and {Winn}, Joshua N.},
        title = "{TESS Eclipsing Binary Stars. I. Short-cadence Observations of 4584 Eclipsing Binaries in Sectors 1-26}",
      journal = {\apjs},
         year = 2022,
        month = jan,
       volume = {258},
       number = {1},
          eid = {16},
        pages = {16},
          doi = {10.3847/1538-4365/ac324a},
archivePrefix = {arXiv},
       eprint = {2110.13382},
 primaryClass = {astro-ph.SR},
       adsurl = {https://ui.adsabs.harvard.edu/abs/2022ApJS..258...16P}
}

@ARTICLE{1978ApJ...223..252B,
       author = {{Bond}, H.~E. and {Liller}, W. and {Mannery}, E.~J.},
        title = "{UU Sagittae: eclipsing nucleus of the planetary nebula Abell 63.}",
      journal = {\apj},
         year = 1978,
        month = jul,
       volume = {223},
        pages = {252-259},
          doi = {10.1086/156257},
       adsurl = {https://ui.adsabs.harvard.edu/abs/1978ApJ...223..252B}
}

@ARTICLE{2022MNRAS.513.2437P,
       author = {{Penoyre}, Zephyr and {Belokurov}, Vasily and {Evans}, N. Wyn},
        title = "{Astrometric identification of nearby binary stars - I. Predicted astrometric signals}",
      journal = {\mnras},
         year = 2022,
        month = jun,
       volume = {513},
       number = {2},
        pages = {2437-2456},
          doi = {10.1093/mnras/stac959},
archivePrefix = {arXiv},
       eprint = {2111.10380},
 primaryClass = {astro-ph.SR},
       adsurl = {https://ui.adsabs.harvard.edu/abs/2022MNRAS.513.2437P}
}

@ARTICLE{2009AJ....138..466H,
       author = {{Hoffman}, D.~I. and {Harrison}, T.~E. and {McNamara}, B.~J.},
        title = "{Automated Variable Star Classification Using the Northern Sky Variability Survey}",
      journal = {\aj},
         year = 2009,
        month = aug,
       volume = {138},
       number = {2},
        pages = {466-477},
          doi = {10.1088/0004-6256/138/2/466},
       adsurl = {https://ui.adsabs.harvard.edu/abs/2009AJ....138..466H}
}

@ARTICLE{1968BAICz..19...90K,
       author = {{Kromov}, G.~S. and {Kohoutek}, L.},
        title = "{Morphological study of planetary nebulae. III. Unclassified and peculiar objects}",
      journal = {Bulletin of the Astronomical Institutes of Czechoslovakia},
         year = 1968,
        month = jan,
       volume = {19},
        pages = {90},
       adsurl = {https://ui.adsabs.harvard.edu/abs/1968BAICz..19...90K}
}

@ARTICLE{2012A&A...548A.109Z,
       author = {{Ziegler}, M. and {Rauch}, T. and {Werner}, K. and {K{\"o}ppen}, J. and {Kruk}, J.~W.},
        title = "{BD-22{\textdegree}3467, a DAO-type star exciting the nebula Abell 35}",
      journal = {\aap},
         year = 2012,
        month = dec,
       volume = {548},
          eid = {A109},
        pages = {A109},
          doi = {10.1051/0004-6361/201219536},
archivePrefix = {arXiv},
       eprint = {1210.7614},
 primaryClass = {astro-ph.SR},
       adsurl = {https://ui.adsabs.harvard.edu/abs/2012A&A...548A.109Z}
}

@ARTICLE{2010AJ....140.1868W,
       author = {{Wright}, Edward L. and {Eisenhardt}, Peter R.~M. and {Mainzer}, Amy K. and {Ressler}, Michael E. and {Cutri}, Roc M. and {Jarrett}, Thomas and {Kirkpatrick}, J. Davy and {Padgett}, Deborah and {McMillan}, Robert S. and {Skrutskie}, Michael and {Stanford}, S.~A. and {Cohen}, Martin and {Walker}, Russell G. and {Mather}, John C. and {Leisawitz}, David and {Gautier}, III, Thomas N. and {McLean}, Ian and {Benford}, Dominic and {Lonsdale}, Carol J. and {Blain}, Andrew and {Mendez}, Bryan and {Irace}, William R. and {Duval}, Valerie and {Liu}, Fengchuan and {Royer}, Don and {Heinrichsen}, Ingolf and {Howard}, Joan and {Shannon}, Mark and {Kendall}, Martha and {Walsh}, Amy L. and {Larsen}, Mark and {Cardon}, Joel G. and {Schick}, Scott and {Schwalm}, Mark and {Abid}, Mohamed and {Fabinsky}, Beth and {Naes}, Larry and {Tsai}, Chao-Wei},
        title = "{The Wide-field Infrared Survey Explorer (WISE): Mission Description and Initial On-orbit Performance}",
      journal = {\aj},
         year = 2010,
        month = dec,
       volume = {140},
       number = {6},
        pages = {1868-1881},
          doi = {10.1088/0004-6256/140/6/1868},
archivePrefix = {arXiv},
       eprint = {1008.0031},
 primaryClass = {astro-ph.IM},
       adsurl = {https://ui.adsabs.harvard.edu/abs/2010AJ....140.1868W}
}

@ARTICLE{2013MNRAS.436.2082T,
       author = {{Tyndall}, A.~A. and {Jones}, D. and {Boffin}, H.~M.~J. and {Miszalski}, B. and {Faedi}, F. and {Lloyd}, M. and {Boumis}, P. and {L{\'o}pez}, J.~A. and {Martell}, S. and {Pollacco}, D. and {Santander-Garc{\'\i}a}, M.},
        title = "{Two rings but no fellowship: LoTr 1 and its relation to planetary nebulae possessing barium central stars}",
      journal = {\mnras},
         year = 2013,
        month = dec,
       volume = {436},
       number = {3},
        pages = {2082-2095},
          doi = {10.1093/mnras/stt1713},
archivePrefix = {arXiv},
       eprint = {1309.4307},
 primaryClass = {astro-ph.SR},
       adsurl = {https://ui.adsabs.harvard.edu/abs/2013MNRAS.436.2082T}
}

@ARTICLE{2021MNRAS.506.4151M,
       author = {{Merc}, J. and {G{\'a}lis}, R. and {Wolf}, M. and {Velez}, P. and {Buil}, C. and {Sims}, F. and {Bohlsen}, T. and {Vra{\v{s}}{\v{t}}{\'a}k}, M. and {Boussin}, C. and {Boussier}, H. and {Cazzato}, P. and {Diarrasouba}, I. and {Teyssier}, F.},
        title = "{Spectroscopic and photometric analysis of symbiotic candidates - I. Ten candidates on classical symbiotic stars}",
      journal = {\mnras},
         year = 2021,
        month = sep,
       volume = {506},
       number = {3},
        pages = {4151-4162},
          doi = {10.1093/mnras/stab2034},
archivePrefix = {arXiv},
       eprint = {2107.05953},
 primaryClass = {astro-ph.SR},
       adsurl = {https://ui.adsabs.harvard.edu/abs/2021MNRAS.506.4151M}
}

@ARTICLE{frew11,
       author = {{Frew}, David J. and {Stanger}, Jeff and {Fitzgerald}, Michael and {Parker}, Quentin and {Danaia}, Lena and {McKinnon}, David and {Guerrero}, Mart{\'\i}n A. and {Hedberg}, John and {Hollow}, Robert and {An}, Yvonne and {Bor}, Shu Han and {Colman}, Isabel and {Graham-White}, Claire and {Li}, Qing Wen and {Mai}, Juliette and {Papadakis}, Katerina and {Picone-Murray}, Julia and {Hoang}, Melanie Vo and {Yean}, Vivian},
        title = "{K 1-6: An Asymmetric Planetary Nebula with a Binary Central Star}",
      journal = {\pasa},
         year = 2011,
        month = mar,
       volume = {28},
       number = {1},
        pages = {83-94},
          doi = {10.1071/AS10017},
archivePrefix = {arXiv},
       eprint = {1009.5914},
 primaryClass = {astro-ph.SR},
       adsurl = {https://ui.adsabs.harvard.edu/abs/2011PASA...28...83F}
}

@ARTICLE{adam14,
       author = {{Adam}, C. and {Mugrauer}, M.},
        title = "{HIP 3678: a hierarchical triple stellar system in the centre of the planetary nebula NGC 246}",
      journal = {\mnras},
         year = 2014,
        month = nov,
       volume = {444},
       number = {4},
        pages = {3459-3465},
          doi = {10.1093/mnras/stu1677},
archivePrefix = {arXiv},
       eprint = {1409.5339},
 primaryClass = {astro-ph.SR},
       adsurl = {https://ui.adsabs.harvard.edu/abs/2014MNRAS.444.3459A}
}

@ARTICLE{miszalski19,
       author = {{Miszalski}, Brent and {Manick}, Rajeev and {Rauch}, Thomas and {I{\l}kiewicz}, Krystian and {Van Winckel}, Hans and {Miko{\l}ajewska}, Joanna},
        title = "{Two's company, three's a crowd: SALT reveals the likely triple nature of the nucleus of the extreme abundance discrepancy factor planetary nebula Sp 3}",
      journal = {\pasa},
         year = 2019,
        month = nov,
       volume = {36},
          eid = {e042},
        pages = {e042},
          doi = {10.1017/pasa.2019.36},
archivePrefix = {arXiv},
       eprint = {1908.08724},
 primaryClass = {astro-ph.SR},
       adsurl = {https://ui.adsabs.harvard.edu/abs/2019PASA...36...42M}
}

@ARTICLE{jones19,
       author = {{Jones}, David and {Pejcha}, Ond{\v{r}}ej and {Corradi}, Romano L.~M.},
        title = "{On the triple-star origin of the planetary nebula Sh 2-71}",
      journal = {\mnras},
         year = 2019,
        month = oct,
       volume = {489},
       number = {2},
        pages = {2195-2203},
          doi = {10.1093/mnras/stz2293},
archivePrefix = {arXiv},
       eprint = {1908.04582},
 primaryClass = {astro-ph.SR},
       adsurl = {https://ui.adsabs.harvard.edu/abs/2019MNRAS.489.2195J}
}

@ARTICLE{ciardullo99,
       author = {{Ciardullo}, Robin and {Bond}, Howard E. and {Sipior}, Michael S. and {Fullton}, Laura K. and {Zhang}, C. -Y. and {Schaefer}, Karen G.},
        title = "{A HUBBLE SPACE TELESCOPE Survey for Resolved Companions of Planetary Nebula Nuclei}",
      journal = {\aj},
         year = 1999,
        month = jul,
       volume = {118},
       number = {1},
        pages = {488-508},
          doi = {10.1086/300940},
archivePrefix = {arXiv},
       eprint = {astro-ph/9904043},
 primaryClass = {astro-ph},
       adsurl = {https://ui.adsabs.harvard.edu/abs/1999AJ....118..488C}
}

@ARTICLE{fies,
       author = {{Telting}, J.~H. and {Avila}, G. and {Buchhave}, L. and {Frandsen}, S. and {Gandolfi}, D. and {Lindberg}, B. and {Stempels}, H.~C. and {Prins}, S. and {NOT staff}},
        title = "{FIES: The high-resolution Fiber-fed Echelle Spectrograph at the Nordic Optical Telescope}",
      journal = {Astronomische Nachrichten},
         year = 2014,
        month = jan,
       volume = {335},
       number = {1},
        pages = {41},
          doi = {10.1002/asna.201312007},
       adsurl = {https://ui.adsabs.harvard.edu/abs/2014AN....335...41T}
}

@INPROCEEDINGS{HarpsN,
       author = {{Cosentino}, Rosario and {Lovis}, Christophe and {Pepe}, Francesco and {Collier Cameron}, Andrew and {Latham}, David W. and {Molinari}, Emilio and {Udry}, Stephane and {Bezawada}, Naidu and {Black}, Martin and {Born}, Andy and {Buchschacher}, Nicolas and {Charbonneau}, Dave and {Figueira}, Pedro and {Fleury}, Michel and {Galli}, Alberto and {Gallie}, Angus and {Gao}, Xiaofeng and {Ghedina}, Adriano and {Gonzalez}, Carlos and {Gonzalez}, Manuel and {Guerra}, Jose and {Henry}, David and {Horne}, Keith and {Hughes}, Ian and {Kelly}, Dennis and {Lodi}, Marcello and {Lunney}, David and {Maire}, Charles and {Mayor}, Michel and {Micela}, Giusi and {Ordway}, Mark P. and {Peacock}, John and {Phillips}, David and {Piotto}, Giampaolo and {Pollacco}, Don and {Queloz}, Didier and {Rice}, Ken and {Riverol}, Carlos and {Riverol}, Luis and {San Juan}, Jose and {Sasselov}, Dimitar and {Segransan}, Damien and {Sozzetti}, Alessandro and {Sosnowska}, Danuta and {Stobie}, Brian and {Szentgyorgyi}, Andrew and {Vick}, Andy and {Weber}, Luc},
        title = "{Harps-N: the new planet hunter at TNG}",
    booktitle = {Ground-based and Airborne Instrumentation for Astronomy IV},
         year = 2012,
       editor = {{McLean}, Ian S. and {Ramsay}, Suzanne K. and {Takami}, Hideki},
       series = {Society of Photo-Optical Instrumentation Engineers (SPIE) Conference Series},
       volume = {8446},
        month = sep,
          eid = {84461V},
        pages = {84461V},
          doi = {10.1117/12.925738},
       adsurl = {https://ui.adsabs.harvard.edu/abs/2012SPIE.8446E..1VC}
}

@ARTICLE{HarpsN-DRS,
       author = {{Dumusque}, X. and {Cretignier}, M. and {Sosnowska}, D. and {Buchschacher}, N. and {Lovis}, C. and {Phillips}, D.~F. and {Pepe}, F. and {Alesina}, F. and {Buchhave}, L.~A. and {Burnier}, J. and {Cecconi}, M. and {Cegla}, H.~M. and {Cloutier}, R. and {Collier Cameron}, A. and {Cosentino}, R. and {Ghedina}, A. and {Gonz{\'a}lez}, M. and {Haywood}, R.~D. and {Latham}, D.~W. and {Lodi}, M. and {L{\'o}pez-Morales}, M. and {Maldonado}, J. and {Malavolta}, L. and {Micela}, G. and {Molinari}, E. and {Mortier}, A. and {P{\'e}rez Ventura}, H. and {Pinamonti}, M. and {Poretti}, E. and {Rice}, K. and {Riverol}, L. and {Riverol}, C. and {San Juan}, J. and {S{\'e}gransan}, D. and {Sozzetti}, A. and {Thompson}, S.~J. and {Udry}, S. and {Wilson}, T.~G.},
        title = "{Three years of HARPS-N high-resolution spectroscopy and precise radial velocity data for the Sun}",
      journal = {\aap},
         year = 2021,
        month = apr,
       volume = {648},
          eid = {A103},
        pages = {A103},
          doi = {10.1051/0004-6361/202039350},
archivePrefix = {arXiv},
       eprint = {2009.01945},
 primaryClass = {astro-ph.SR},
       adsurl = {https://ui.adsabs.harvard.edu/abs/2021A&A...648A.103D}
}

@INPROCEEDINGS{not,
       author = {{Djupvik}, Anlaug Amanda and {Andersen}, Johannes},
        title = "{The Nordic Optical Telescope}",
    booktitle = {Highlights of Spanish Astrophysics V},
         year = 2010,
       series = {Astrophysics and Space Science Proceedings},
       volume = {14},
        month = jan,
        pages = {211},
          doi = {10.1007/978-3-642-11250-8_21},
archivePrefix = {arXiv},
       eprint = {0901.4015},
 primaryClass = {astro-ph.IM},
       adsurl = {https://ui.adsabs.harvard.edu/abs/2010ASSP...14..211D}
}

@ARTICLE{jones22,
       author = {{Jones}, David and {Boffin}, Henri M.~J. and {Brown}, Alex J. and {Zak}, Jiri and {Hume}, George and {Munday}, James and {Miszalski}, Brent},
        title = "{A detailed study of the barium central star of the planetary nebula Abell 70}",
      journal = {\mnras},
         year = 2022,
        month = nov,
       volume = {516},
       number = {4},
        pages = {4833-4843},
          doi = {10.1093/mnras/stac2501},
archivePrefix = {arXiv},
       eprint = {2208.14778},
 primaryClass = {astro-ph.SR},
       adsurl = {https://ui.adsabs.harvard.edu/abs/2022MNRAS.516.4833J}
}

@ARTICLE{miszalski13,
       author = {{Miszalski}, B. and {Boffin}, H.~M.~J. and {Jones}, D. and {Karakas}, A.~I. and {K{\"o}ppen}, J. and {Tyndall}, A.~A. and {Mohamed}, S.~S. and {Rodr{\'\i}guez-Gil}, P. and {Santander-Garc{\'\i}a}, M.},
        title = "{SALT reveals the barium central star of the planetary nebula Hen 2-39}",
      journal = {\mnras},
         year = 2013,
        month = dec,
       volume = {436},
       number = {4},
        pages = {3068-3081},
          doi = {10.1093/mnras/stt1795},
archivePrefix = {arXiv},
       eprint = {1309.5239},
 primaryClass = {astro-ph.SR},
       adsurl = {https://ui.adsabs.harvard.edu/abs/2013MNRAS.436.3068M}
}

@ARTICLE{qian24,
       author = {{Qian}, S. -B. and {Zhu}, L. -Y. and {Li}, F. -X. and {Li}, L. -J. and {Han}, Z. -T. and {He}, J. -J. and {Zang}, L. and {Chang}, L. -F. and {Sun}, Q. -B. and {Li}, M. -Y. and {Zhang}, H. -T. and {Yan}, F. -Z.},
        title = "{A Brown Dwarf Orbiting around the Planetary-nebula Central Binary KV Vel}",
      journal = {\apj},
         year = 2024,
        month = sep,
       volume = {972},
       number = {1},
          eid = {13},
        pages = {13},
          doi = {10.3847/1538-4357/ad631a},
archivePrefix = {arXiv},
       eprint = {2409.02480},
 primaryClass = {astro-ph.SR},
       adsurl = {https://ui.adsabs.harvard.edu/abs/2024ApJ...972...13Q}
}

\begin{appendix}

\begin{figure*}
\section{Additional figures}

\centering
\includegraphics[width=0.9\textwidth]{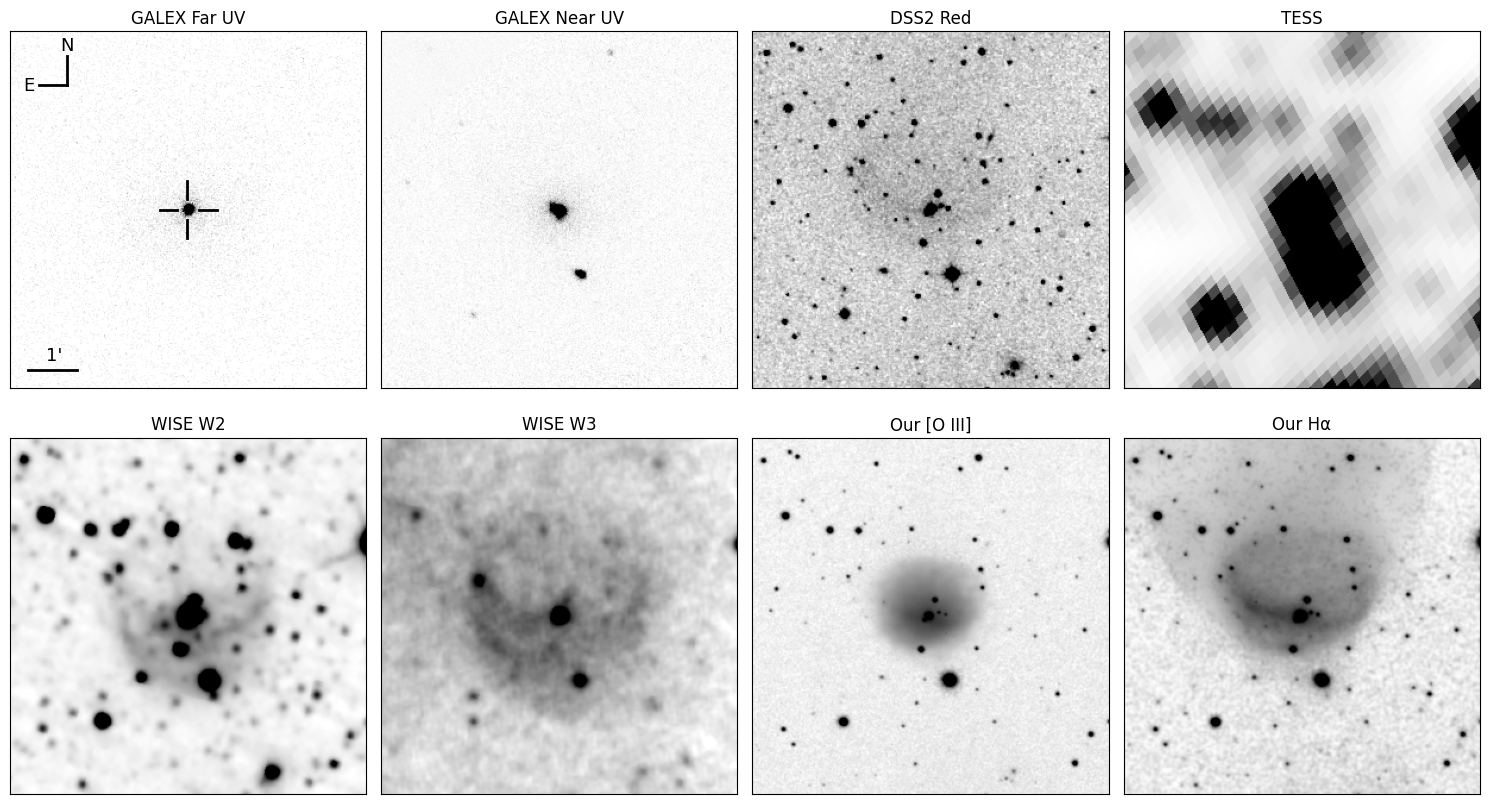}
\caption{A 7.2\arcmin{} field centred on the central star of K~1-6 (marked by a cross), shown in \textit{GALEX} Far-UV, \textit{GALEX} Near-UV \citep{2005ApJ...619L...1M}, DSS2 Red, and TESS images (upper row), alongside WISE W2 and WISE W3 observations, as well as our [\ion{O}{iii}] and H$\alpha$ imaging (bottom row).}
\label{fig:mosaic}
\end{figure*}

\begin{figure*}
\centering
\includegraphics[width=\textwidth]{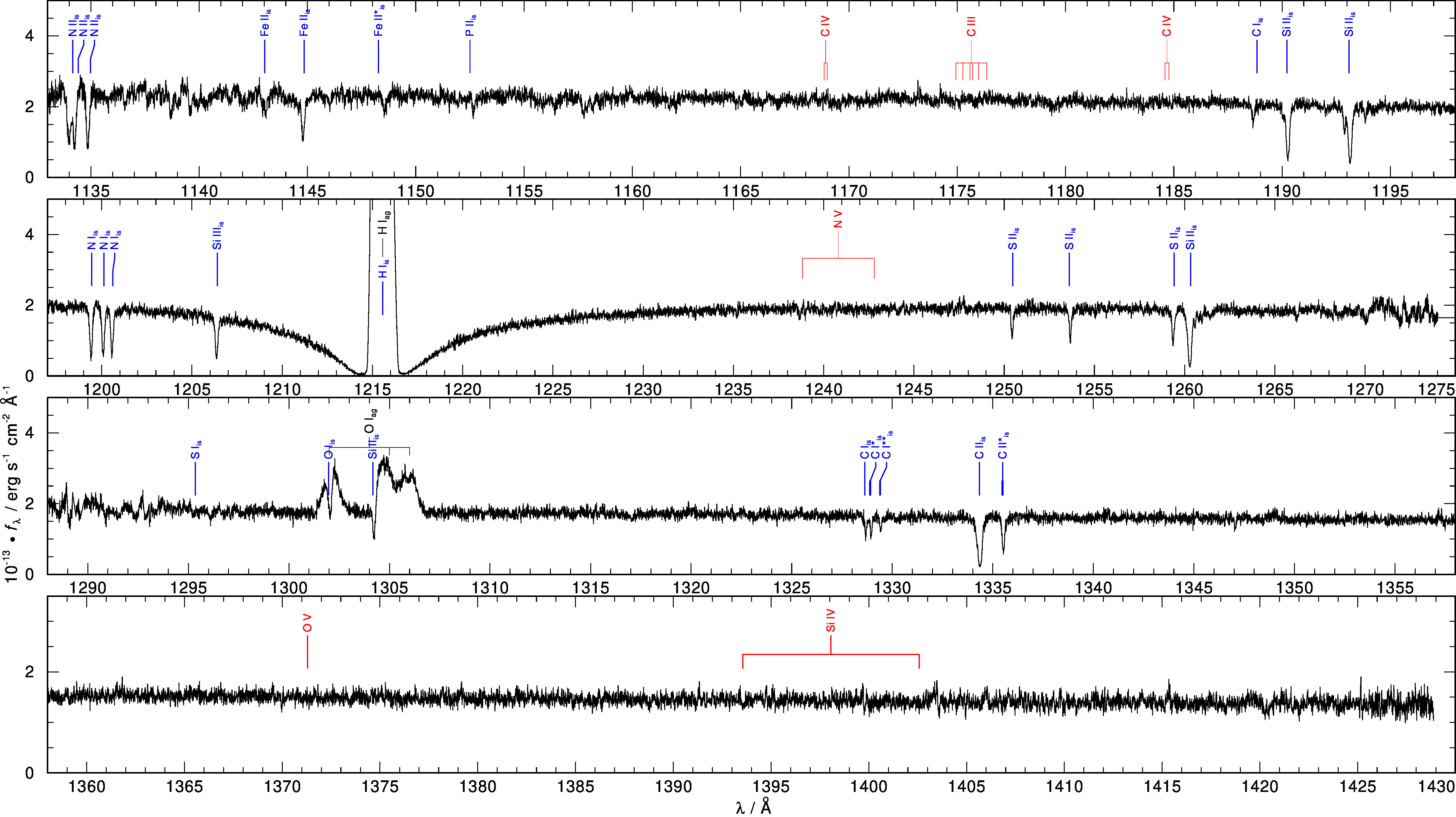}
\caption{HST/COS G130M spectroscopy of the white dwarf. The locations of typically strong photospheric lines are highlighted in red, interstellar and airglow lines are highlighted in blue and black, respectively.}
\label{fig:hst1}
\end{figure*}
\clearpage
\begin{figure}
\centering
\includegraphics[width=\columnwidth]{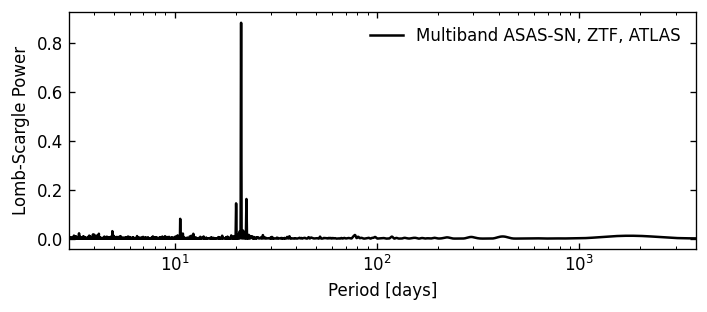}
\includegraphics[width=\columnwidth]{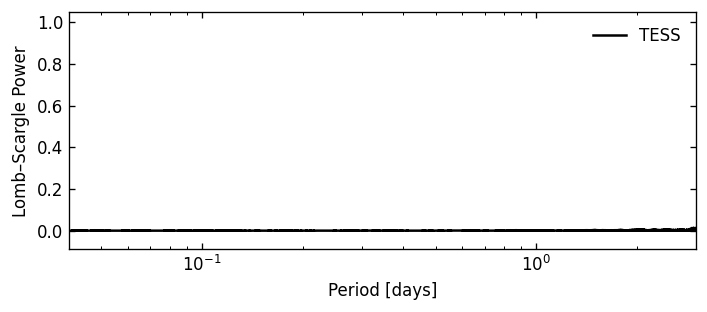}
\caption{Periodograms of the K~1-6 light curves. Upper panel: Multiband Lomb--Scargle periodogram computed from the ASAS-SN \textit{V} and \textit{g}, ZTF \textit{g}, and ATLAS \textit{o} light curves over periods from 3 days to approximately $0.8\times$ the total time baseline. The dominant 21.3-day period, its first harmonic ($P/2$), and the expected one-year alias peaks are visible. The periodograms of the individual bands are nearly identical. Lower panel: Lomb--Scargle periodogram of the TESS light curve after removal of the long-term modulation, computed over periods between 1 hour and 3 days.}
\label{fig:periodograms}
\end{figure}

\section{On the proposed tertiary component of Abell 63}\label{app:a63}
The possible existence of a tertiary component in the planetary nebula Abell~63 was proposed by \citet{ciardullo99}, based on the projected proximity of an additional star to the central binary in \textit{Hubble Space Telescope} (\textit{HST}) images obtained in 1993. The central star of Abell~63, known also as UU~Sge, is an eclipsing binary system with an orbital period of 11.16 hours \citep{1978ApJ...223..252B,2022ApJS..258...16P}.

\begin{figure}
\centering
\includegraphics[width=\columnwidth]{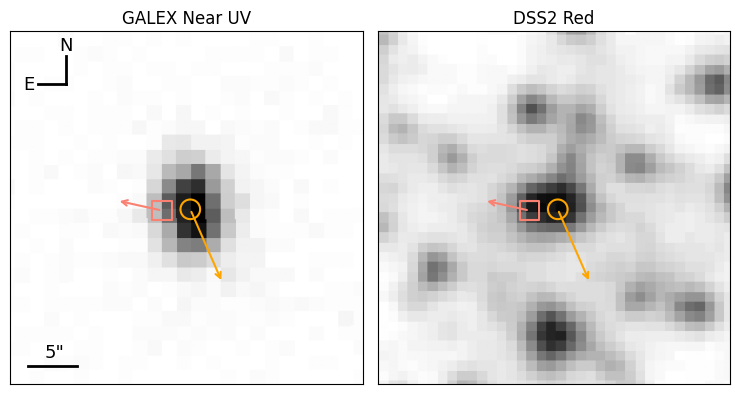}
\caption{The 36\arcsec{} field around the central star of Abell 63 (indicated by the orange circle). The position of the proposed tertiary companion is marked by the pink square. Proper motion vectors for both objects are shown as arrows. The left panel displays the \textit{GALEX} Near-UV image, while the right panel shows the DSS2 Red image.
}
\label{fig:abell63}
\end{figure}

\citet{ciardullo99} reported a separation of 2.82\arcsec{} and a position angle of 94\textdegree{} between UU~Sge and the nearby object. These two sources are also resolved in \textit{Gaia}~DR3, which reports a slightly larger separation of 2.88\arcsec{} at epoch 2016.0 (see Fig.~\ref{fig:abell63}).

However, the astrometric data from \textit{Gaia} suggest that the two stars are not physically associated. The parallax of the central binary is $0.3651 \pm 0.0261$~mas, corresponding to a distance of approximately 2.7~kpc, while the nearby star has a parallax of $0.5617 \pm 0.0345$~mas, indicating a closer distance of about 1.8~kpc. Their proper motions also differ significantly. The central binary has proper motions of $-3.2775 \pm 0.0242$~mas~yr$^{-1}$ in right ascension and $-7.497 \pm 0.0223$~mas~yr$^{-1}$ in declination, whereas the nearby star shows values of $4.621 \pm 0.0299$~mas~yr$^{-1}$ and $1.0282 \pm 0.0288$~mas~yr$^{-1}$, respectively. Neither star has a radial velocity measurement in the \textit{Gaia} database.

Based on the discrepancy in parallaxes and proper motions, we conclude that the projected companion is most likely a chance alignment and not physically bound to the Abell~63 system.

\section{Line profile variations in the spotted model}\label{app:line_profiles}

\begin{figure}
\centering
\includegraphics[width=\columnwidth]{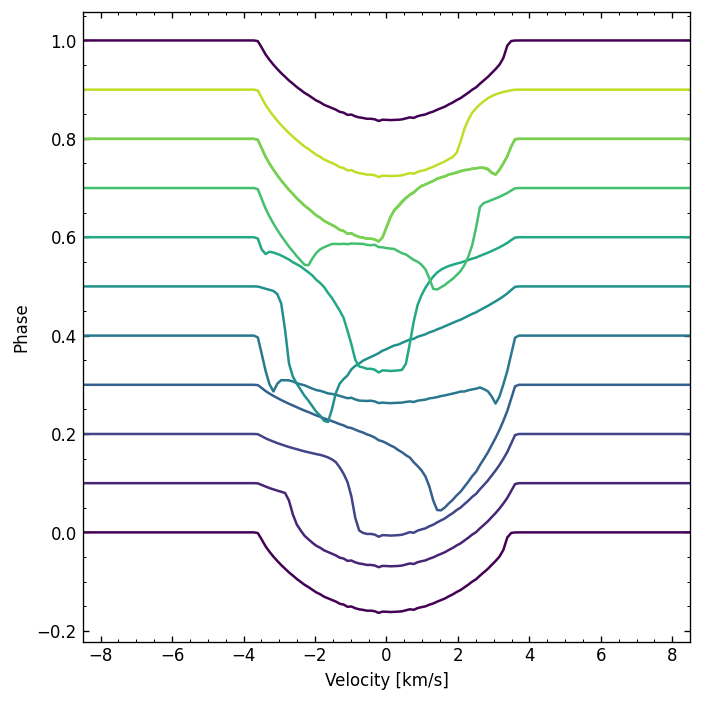}
\caption{Predicted variations of spectral line profiles as a function of photometric phase from the spotted-star model. The model assumes a single K2 star with a radius of 1.5\,\Rsol with two large cool spots.}
\label{fig:line_profiles}
\end{figure}

To interpret the photometric variability of K~1-6, we explored a rotating spotted-star model using \textsc{phoebe2} \citep{phoebe2,phoebe3,phoebe4,phoebe5} as outlined in Sect.~\ref{sec:rotation}. In addition to reproducing the light-curve morphology, the model predicts phase-dependent distortions of spectral line profiles (Fig.~\ref{fig:line_profiles}).

These distortions arise from the non-uniform surface brightness distribution caused by large, cool spots and can lead to asymmetric or multi-peaked line profiles. We suggest that this effect is responsible for the apparent double-peaked CCFs observed in some of our spectra.

\end{appendix}

\end{document}